\documentclass[a4paper,11pt]{article}
\pdfoutput=1 

\usepackage{jcappub} 

\usepackage[T1]{fontenc} 
\usepackage{siunitx}

\title{\boldmath  Cosmological implications of  electromagnetically interacting dark matter: milli-charged particles and atoms with singly and doubly charged dark matter}

\author[a]{Gautham A P}
\author[b]{Shiv Sethi}

\affiliation[a]{Indian Institute of Technology Madras(IITM),Chennai, India}
\affiliation[b]{Raman Research Institute(RRI),Bangalore, India}

\emailAdd{apgautham@gmail.com, sethi@rri.res.in}

\abstract{While the behavior  of the dominant component of the dark matter is reasonably well established by cosmological
observables, its particle nature  and interactions with the rest of the matter are not known. 
We consider three dark matter  models that admit electromagnetic interaction between
baryons and  dark matter: (a) milli-charged particle (CCDM) of charge $q_{\rm ccdm}$ and mass $m_{\rm ccdm}$, (b) a neutral  atom of two charged particles of  mass $m_{\rm dd}$ (DD), and (c) a neutral  atom of doubly charged particle and helium nucleus (HeD). 
We derive and discuss in detail the formation, stability, and interaction of these atoms with baryons. We derive the implications of this new interaction in the tight-coupling approximation, which allows us to analytically
gauge their impact on the matter power spectrum and CMB anisotropy. We  incorporate this new interaction  into  the publicly-available code CLASS to obtain numerical results. We compare our results with Planck 2018 data    to  constrain the fraction of interacting 
dark matter. For the range of allowed astrophysical parameters, the HeD atom yields the results of $\Lambda$CDM model for $k < 1 \, \rm Mpc^{-1}$, and hence its fraction is not constrained by CMB anisotropy data which is sensitive to $k < 0.2 \, \rm Mpc^{-1}$. For  $m_{\rm dd} \gtrsim 25 \, \rm GeV$, the DD atom is also not constrained by CMB data.  For $m_{\rm dd} = 10 \, \rm GeV$, the CMB data constrains the fraction of DD atoms to be smaller than
4\% of the total CDM component. For   $q_{\rm ccdm} = 10^{-6}e$ and $m_{\rm ccdm} = 50 \, \rm MeV$, the CCDM fraction is
constrained to be less than 1\%. }

\begin{document}
\maketitle
\flushbottom

\newpage

\section{Introduction}
\label{sec:Introduction}

The nature of dark matter remains a mystery even though the existence of this  
component  has been  well established by many  observations covering a wide range of mass  scales---from dwarf galaxies to clusters of galaxies--- 
and different epochs of the universe through cosmological probes. This list includes  cosmic microwave background (CMB) anisotropy experiments \cite{Planck2018,Ade:2015xua,Hinshaw:2012aka,Sievers:2013ica}, large scale structure surveys \cite{Beutler:2016ixs,Tegmark:2006az,Tegmark:2003ud}, the study of the galaxy rotation 
curves \cite{Begeman:1991iy}, cosmological weak gravitational lensing observations \cite{Bartelmann:1999yn,Clowe:2006eq}, etc. 

 The Weakly Interacting Massive Particle (WIMP) is normally considered the 
leading candidate for the  cold dark matter (CDM) in the universe. Its
popularity is partly   inspired by the 
well-known WIMP miracle\cite{Craig:2015xla}. The supersymmetric extension of the standard model of particle physics allows for a stable  particle in the mass range $10\hbox{--}1000 \, \rm GeV$  with self-annihilation cross-section $\langle \sigma v\rangle \simeq  10^{-26}\rm{cm^3s^{-1}}$; these parameters predict
 the   abundance of cold dark matter in the observed range. This theoretical insight
has inspired many direct, indirect and collider searches of the WIMP  \cite{Angloher:2011uu,Aprile:2010um,Ahmed:2010wy,Adriani:2010rc,FermiLAT:2011ab,Barwick:1997ig,Aguilar:2007yf,Goodman:2010yf,Fox:2011pm}.  Even though these
searches  have not  yet succeeded, the sensitivity of the direct detection
experiments, LUX and XENON, has improved considerably in 
the recent years \cite{2018PhRvL.121k1302A,2017PhRvL.118y1302A}. 

The cold dark matter paradigm based on WIMP has  long-standing astrophysical  issues: core-cusp problem \cite{deBlok:2009sp,Navarro:1995iw,Walker:2011zu}, missing satellites of the Milky way \cite{Klypin:1999uc,Moore:1999nt}, ``too big to fail'' issue
\cite{Garrison-Kimmel:2014vqa,BoylanKolchin:2011de}.  All these issues probably indicate a need  to go beyond the  WIMP model  and to consider alternative candidates, which differ from the model  on galactic scales but must reproduce its success on 
cosmological scales.
The focus of this paper is to study cosmological implications of dark matter
particles with electromagnetic interactions. Such models were proposed 
 before precision cosmological data became available \cite{1990NuPhB.333..173D,1990PhRvD..41.2388D}. In such models, charged particles that survive annihilation in the early universe could form atoms with protons and helium nuclei and provide the observed mass density of cold  dark matter in the universe. Many such models were ruled out 
and revised as better astrophysical and  cosmological data became available (e.g. \cite{2013PhRvD..87j3515C,2012PhRvD..85j1302C,2017AIPC.1900d0002B,2010PhLB..682..337K,2009JCAP...07..014C,Dvorkin:2013cea,Xu:2018efh}). The current  cosmological data clearly rules out 
a dark matter particle with electronic charge unless it is very heavy. One possible extension of this idea 
was a neutral  atom formed with  the proton and a negatively charged dark matter
particle (e.g. \cite{1990NuPhB.333..173D}). Such an atom is not stable to charge-exchange reaction and therefore
does not remain neutral as the universe evolves (for details of charge exchange
reactions see e.g. \cite{2011piim.book.....D}). 

We consider three models for our study. The first model  posits the presence of 
a milli-charged dark matter particle (CCDM model, e.g. \cite{Dubovsky:2003yn,2012PhRvD..85j1302C,Kovetz:2018zan}). The  model---parameterized by the magnitude of milli-charge and the mass of the dark matter particle--- has been extensively studied against cosmological data (e.g. \cite{Kovetz:2018zan}). The second model
assumes the presence of two dark matter particles of positive and negative electronic charge which
recombine to form a neutral  atom (DD model). The stability of   this atom to charge exchange reaction  constrain the masses of charged particles. The third model, which is a natural extension of the proton-dark matter atom, is based on a doubly charged dark matter particle (HeD model, see \cite{2012AIPC.1446...99K} for theoretical motivation for such a scenario)  which recombines with a helium nucleus to form a neutral atom. The mass of this particle must be much larger
than the helium nucleus to ensure, (a) only a small fraction of helium nuclei recombine with   the  dark matter particle to satisfy nucleosynthesis bounds, and (b) the atom is stable to charge-exchange reactions. We discuss in the detail
the recombination, ionization, and  charge-exchange stability of such  atoms, in addition to computing the scattering of such atoms off baryons.

One of the motivation of studying such models is that  these models might  allow, for some parameter range,    smaller matter power at small scales as compared to the WIMP model. So some of these models can be viewed as complementary  to other existing models to 
explain the missing small scale power (e.g. warm dark matter (WDM) model, \cite{Dolgov:2000ew, Viel:2005qj} and other models such as LFDM, CHDM, or ULA;  for details see e.g. 
\cite{Sigurdson:2003vy,2017PhRvD..95d3541H,Sarkar:2015dib})
The milli-charged dark matter model has recently been invoked to explain the 
EDGES results \cite{EDGES2018,Barkana2018,fialkov18,Kovetz:2018zan}.

In the next  section,  we discuss the three dark matter
  models we consider and   study their formation, stability,  and interaction
  with baryons. In section~\ref{sec:evol_cou}, we consider the  evolution of coupled baryon-dark matter
  system in linear perturbation theory relevant for the early universe.
  In particular, we show how the new interaction can be incorporated within
  the framework of tight-coupling approximation, which allows us to gauge
  the impact of the new interaction on cosmological observables. We also discuss how this approximation allows us to numerically implement the altered evolution of coupled fluids. 
In section~\ref{sec:Compare}, we study  the impact of the  new interaction on   cosmological observables such as the matter power spectrum and CMB anisotropies  and also  provide  results of  MCMC analysis which assesses  the viability of 
the new models against  the  Planck data.  In section~\ref{sec:disc} we summarize and discuss our results.

For computing the matter power spectra for the  proposed models, 
we assume a spatially flat universe with its   best-fit 
cosmological parameters estimated by Planck \cite{Planck2018}: $\Omega_b = 0.049$,   $\Omega_{\rm dm} = 0.254$ (this corresponds to the total cold dark matter  content of the universe in our case as we consider models where there are two components of the dark matter), and $h= 0.67$. 

\section{Dark matter models with electromagnetic interaction} \label{sec:appendix}
In this section we discuss in detail the formation of atoms with charged dark matter and their interaction with baryons. 
\subsection{Milli-Charged Dark Matter}
\label{sec:Milli-Charged Dark Matter}
The  milli-Charged dark matter  can interact with charged baryons  via coulomb interaction. Assuming baryons and dark matter particles to be at different temperatures and averaging over the thermal distribution, we get the interaction rate to be (e.g. \cite{Kovetz:2018zan}):
\begin{equation}
\frac{1}{\tau_{\rm bd}} = n_{\rm ccdm} \frac{4 \pi}{3} \sqrt{\frac{2}{\pi}} \left(\frac{e q_{\rm ccdm}}{\mu}\right)^{2}\frac{m_{\rm ccdm}}{m_{\rm ccdm}+m}\log_e \Lambda \left [3 k_b \left(\frac{T}{m} + \frac{T_{\rm ccdm}}{m_{\rm ccdm}} \right ) \right]^{-\frac{3}{2}}  
\label{eq:col_int}
\end{equation}
Here $m$ and $T$  stand for the mass and temperature of either electron, proton, or
helium nucleus while $m_{\rm ccdm}$,  $T_{\rm ccdm}$, and $q_{\rm ccdm}$ denote  the mass,
temperature,  and charge, respectively, of the milli-charged dark matter particle. $\mu$ is the reduced mass of $m$ and $m_{ccdm}$ and  $\Lambda$ is the usual coulomb logarithm. If $f_{\rm ccdm}$ is the fraction of  cold dark matter in the form of milli-charged particles,  the mass density and  the number density of CCDM dark matter  are: $\rho_{\rm ccdm} = f_{\rm ccdm} \Omega_{\rm dm} \rho_c$ and $n_{\rm ccdm} = f_{\rm ccdm} \Omega_{\rm dm} \rho_c/m_{\rm ccdm}$.
In the expression above we have assumed the dark matter to be the target. For the reverse reaction, the  time scale:
$\tau_{\rm db} = \tau_{\rm bd}\rho_{\rm ccdm}/\rho_{b}$, as discussed in section~\ref{sec:momloss}.

In deriving Eq.~(\ref{eq:col_int}),  we have neglected the bulk peculiar velocity and, as already noted above,  the dark Matter particles obey Maxwell-Boltzmann distribution. Eq.~(\ref{eq:col_int}) shows that the electron-DM interaction  is 
the dominant scattering process. As the time scale of equilibriation between
electron, proton, and Helium nuclei is  much shorter than the expansion rate, 
they share a common temperature. For the purposes of computing CMB anisotropies, 
this means that when electron-DM interaction time scale is shorter than
the expansion time, the dark matter particles can be assumed to be coupled
to the entire baryonic fluid. 

Eq.~(\ref{eq:col_int}) shows that the  interaction rate (the inverse of interaction time)  falls  as $a^{-3/2}$ while
the expansion rate,  $H$,  drops  as $a^{-2}$ in the radiation-dominated era. Therefore, the  coulomb interaction becomes more important  at later  times; the ratio of the interaction  and expansion rate is given in Eq.~(\ref{eq:ccdmhr}).  At high redshifts, the DM and baryons are decoupled, which means that their initial 
temperatures could be  different. However, in this work, we assume these
temperatures to be the same.

\subsection{Neutral atoms with charged  dark matter}

It is well known that a free dark matter particle of  electronic 
charge    is incompatible with CMB and  large scale clustering  data because its strong interaction  with baryons results in a coupled baryon-photon-dark matter fluid\footnote{If the dark matter particle  is very heavy, this 
constraint can be obviated because the dark matter could remain uncoupled
from baryons up to $z\simeq 1000$; the lower limit
on the mass of such  a dark matter particle can be computed from Eq.~(\ref{eq:col_int}) (see e.g. \cite{Kamada_2017})}.
One possible way to reduce interaction between dark matter and baryons is to 
 consider  cases in which dark matter particles form atoms with either another dark matter particle or with baryonic particles (protons or  Helium nuclei\footnote{Primordial nucleosynthesis also produces a small fraction, $\simeq 10^{-5}$ of baryons, of $^3$He. In this paper, the helium nucleus always refers
to $^4$He, whose mass abundance is  nearly 25\% of baryons.}). 

An important consideration in such cases is that the neutral atom be stable to 
charge-exchange reactions (to be discussed in more detail below). For instance, the dark matter carrying  electronic  charge   can  form a neutral  atom with protons.  However, such an atom is unstable 
to charge-exchange  reaction in which the proton is replaced by the helium nucleus in the atom, yielding an atom with a net charge. 

This motivates us to  consider two cases: (a) a neutral  atom formed with a doubly-charged DM particle of mass $m_{\rm hed}$  with Helium  nucleus such that  $m_{\rm hed} >  m_{\rm he}$ \footnote{As discussed in section~\ref{sec:Compare},
  we require $m_{\rm hed} \gg   m_{\rm he}$ to satisfy nucleosynthesis and CMB
  constraints.} and (b) an atom formed with two singly charged DM particles of masses  $m_1$ and $m_2$ 
such that   $m_1, m_2 \gg  m_{\rm he}$ or both the particles are much heavier than the helium nucleus. The two masses  $m_1$, $m_2$ are otherwise unconstrained 
and we consider the case when $m_1=m_2$; we denote this common mass as  $m_{\rm dd}$. 

\subsubsection{Formation of $\rm HeD$ and $\rm DD$ atoms}

The three most important physical processes for the formation and 
destruction of these atoms are: (a) photoionization, (b) recombination, and
(c) charge exchange. 

The cross section of charge-exchange reaction is 
on the same order of magnitude as the elastic scattering cross section between
the charged particle and the neutral atom, discussed in the next subsection,   if the reaction is exothermic or the resultant atom is more stable (e.g. \cite{2011piim.book.....D}).

Protons and Helium nuclei can form atoms with   negatively-charged  dark matter particles. If dark matter is singly charged then protons can form neutral atoms with
dark matter but charge-exchange reaction replaces proton with helium nucleus.
This reaction is energetically favored as it results in a more stable atom. 
Using the analysis of the next section it can be shown that  the rate of this reaction exceeds the expansion rate for $z > 10^5$ therefore 
it is not possible to 
sustain  a neutral dark matter atom with a single proton. This is our primary 
motivation for considering a doubly charged dark matter particle. 

Two  dark matter  particles carrying electronic  charge can form a stable neutral atom only if  the two equal mass  particles are  much heavier than
the helium nucleus (to be discussed later). If these particles are lighter than the Helium nucleus, 
charge-exchange reaction would replace one of the particles with helium nucleus.

If the dark matter is doubly charged, it could form two different 
neutral atoms: $\rm ppD$ or $\rm HeD$. The charge-exchange 
reaction would turn the first atom into a more stable, $\rm pHeD$, and therefore 
it would not  remain neutral. The only stable neutral atom is $\rm HeD$. Therefore, if the dark matter atom is to remain neutral during the pre-recombination phase, it is imperative that a large fraction of  the dark matter particles are captured to form $\rm HeD$  before $\rm ppD$ could form   during the evolution of the universe.

We compute the rate of formation of hydrogenic atoms $\rm pD$ and $\rm HeD$ to address this issue.  All quantities are scaled with respect to corresponding 
cross sections for the  hydrogen atom. 

{\it Binding energy}: For $m_{\rm dm} \gg  m_{\rm He}$, the binding energies of the 
$\rm pD$ and $\rm HeD$  atoms are $Z^2 (m_p/m_e) \,  \rm  Ry$ and $Z^4 (m_{\rm He^4}/m_e) \, \rm Ry$, respectively.  For the  $\rm DD$ atom, the binding energy is $(m_{\rm dd}/2m_e) \,  \rm  Ry$. Here $1 {\rm Ry} = 13.6 \, \rm eV$ is the binding energy of the hydrogen atom.

{\it Ionization cross section}: The ionization cross section of different atoms
can be computed from appropriate scaling of the cross section for  hydrogen atom. We compute 
these cross sections at the ionization  threshold of each species, where it is the maximum, and express them in terms of hydrogen ionization cross section. 
The ionization cross sections are\footnote{These cross-sections can be 
computed from positive-energy spherically-symmetric solutions of a hydrogenic atom, which are valid for the ionization process close to the threshold; for details see e.g. chapter~IVb of  \cite{1957qmot.book.....B} or section~148, problem~4 of \cite{1965qume.book.....L}.}: 
\begin{eqnarray}
\sigma^{\rm ion}_{\rm pD}(\nu_{\rm pD}) &  = &  {m_e^2 \sigma^{\rm ion}_{\rm H} \over Z^5 m_p^2} \\
\sigma^{\rm ion}_{\rm HeD}(\nu_{\rm HeD}) &  = &  {m_e^2 \sigma^{\rm ion}_{\rm H} \over Z^8 m_{\rm He}^2} \\
\sigma^{\rm ion}_{\rm DD}(\nu_{\rm DD}) &  = &  {4 m_e^2 \sigma^{\rm ion}_{\rm H} \over m_{\rm dd}^2}
\end{eqnarray}
Here $Z=2$  and $\sigma^{\rm ion}_{\rm H} = 6.3\times 10^{-18} \, \rm cm^2$ is the cross
section of hydrogen at ionization  threshold $\nu = 13.6 \, \rm eV$. 

{\it Recombination cross section}: Using Milne relation (e.g see \cite{1979rpa..book.....R} for details),  the recombination 
cross sections can be computed from ionization cross sections. This gives:
\begin{eqnarray}
\sigma^{\rm rec}_{\rm pD}(v) &  \simeq  &  {\chi_{\rm H}^2  \sigma^{\rm ion}_{\rm H} \over Z  m_p^2 c^2 v^2} \\
\sigma^{\rm rec}_{\rm HeD}(v) &  \simeq  &  {\chi_{\rm H}^2  \sigma^{\rm ion}_{\rm H} \over   m_{\rm He}^2 c^2 v^2} \\
\sigma^{\rm rec}_{\rm DD}(v) &  \simeq  &  {2\chi_{\rm H}^2  \sigma^{\rm ion}_{\rm H} \over   m_{\rm dd}^2 c^2 v^2}
\end{eqnarray}
Here $\chi_{\rm H} = 13.6 \, \rm eV$ is the binding energy of hydrogen atom and 
$v$ is the relative velocity of the two recombining particles. We have 
assumed that this velocity is much smaller than the binding energy of respective atoms.  A quantity of 
interest is the recombination coefficient, $\alpha(T) = \langle v \sigma(v) \rangle$, where the average is over thermal distribution of  particles; all the species  are  in thermal equilibrium owing to the initial condition at temperature $T$ and therefore we could replace $m v^2 \simeq 2kT$ for each species. 

The evolution of the ionization state of each species is given by:
\begin{equation}
{dn_i \over dt} = n_{\rm dm} n_{j} \alpha_i(T) -  c n_i\int_{\nu_i}^\infty n_\gamma(\nu)  \sigma_i(\nu) d\nu
\label{eq:ionrec}
\end{equation}
Here $n_i$ refers to the number density of either $\rm pD$, $\rm HeD$ or $\rm DD$ atoms. $n_j$ corresponds
to  the number  density of either protons or   helium nuclei or, for the case of 
DD atom,  the number density of  dark matter particles. $n_{\rm dm}$ is the number density of dark matter particles, which for the case of $DD$ atom is the same as $n_j$.
$\nu_i$, $\sigma_i$, and $\alpha_i$  give   ionization thresholds, ionization cross sections, and recombination coefficients   for  different atoms. 
At redshifts of interest, the main source of photoionization of these atoms 
is background blackbody radiation  which gives the number density of ionizing photons per unit 
frequency, $n_\gamma = B_\nu(T)/(c h\nu)$, where $B_\nu(T)$ is the blackbody  specific intensity at temperature $T$. 

Eq.~(\ref{eq:ionrec}) can be solved numerically. We do not present numerical 
results here but outline the essential outcome of the solution of  the equation. We first consider $\rm pD$ and $\rm HeD$ atoms. Our aim is two-fold: (a)  to show that the recombination rates are large enough in comparison with expansion and 
ionization rates to  allow  the formation of these atoms, (b)  the DM atoms preferably 
recombine with  helium nuclei and not with  protons.  The rate of ionization at any redshift
can be approximated as $B_{\nu_i}(T)\sigma_i(\nu_i)/h$, where $\nu_i$ corresponds
to ionization threshold. We first compare the ionization rates of the two
atoms  at $z\simeq 2\times 10^8$. At this redshift, the ionization threshold 
of $\rm pD$ atom corresponds to the peak of the  Planckian. At the
redshift, the 
ionization rate of $\rm HeD$ is smaller than $\rm pD$ by more than 10~orders of magnitude. While the former rate is smaller than both the expansion rate and
recombination rate, the latter is much larger than both. Therefore, at this redshift, the rates are  favourable to the formation of $\rm HeD$ atom while $\rm pD$ cannot form. We check that the conditions are  such that nearly all the DM particles recombine to form 
$\rm HeD$ around this redshift.  As $\rm HeD$ atom is more stable, this atom 
cannot be further altered by charge-exchange processes. 

As noted above the $\rm DD$ atom is stable only if its binding energy is much larger than other possible atoms that can form ($\rm pD$ and $\rm HeD$). The ratio of binding energy of a singly charged DM particle forming 
an atom with helium nucleus to $\rm DD$ atom is $Z^2 m_{\rm He}/m_{\rm dd}$. This requires
the DM particle to be at least four times heavier than the helium nucleus. We verify that if the ratio of binding energies is   10  then nearly all the dark matter
particles recombine to form $\rm DD$ atom before  the formation of $\rm HeD$ atom. 

We note that the recombination process cannot result in the capture of all the dark matter particles into neutral atoms, as the recombination process  becomes 
inefficient after  a majority of dark matter particles have already recombined
(the first term of Eq.~(\ref{eq:ionrec})). Therefore, this process will always
leave  a tiny fraction of free  charged dark matter particles. The exact fraction of this residual depends on the  parameters of the models and the details  of the recombination process.  If $\Delta\rho_{\rm dm} \ll \rho_b$, this will have negligible effect  on cosmological observables such 
as CMB anisotropies.  We neglect the  impact of this residual charged 
component of dark matter in our study.

\subsubsection{Charged particle scattering off neutral atoms: different approximations}
We consider the scattering of charged particles with a neutral hydrogenic 
atom, of polarizability $\alpha_0 \simeq a^3$, in its ground state, where $a$ is the Bohr radius of the ground state. For the two cases we consider, the Bohr radii are: $a_{\rm hed} = \hbar^2/(Z^2 e^2 m_{\rm he})$($Z=2$) and $a_{\rm dd} = 2\hbar^2/(e^2 m_{\rm dd})$. (the Bohr radius for hydrogen atom: $a_{h} = \hbar^2/(e^2 m_{\rm e})$)

We first consider this  scattering in Born's approximation. This approximation is 
valid if the velocity of the charged particle is large as compared to the velocity of the lighter particle inside  the atom, $ka \gg 1$. As all the species are in thermal equilibrium with each other, the kinetic energy of all the  particles is $\simeq k_{\rm B}T$, this gives $k\simeq \sqrt{mk_{\rm B}T}/\hbar$ For the two 
cases we consider: 
\begin{eqnarray}
k a_{\rm hed} &\simeq & 2 \times 10^{-3} \left ({T\over 10^6 \, \rm K}  \right)^{1/2} \\
k a_{\rm dd} & \simeq  & 4 \times 10^{-3}  \left ({m_{\rm he} \over m_{\rm dd}} \right )\left ({T\over 10^6 \, \rm K}\right)^{1/2}
\label{eq:bornapp}
\end{eqnarray}
Therefore, except at very high temperatures, Born's approximation in not 
applicable  to  our analysis. 

Another approximation commonly used in  proton scattering off neutral 
atoms in interstellar medium is the semi-classical approximation which treats proton trajectory 
as classical. In this case, the cross section of scattering is $\simeq b_0^2$
where $b_0 \simeq (e^2 \alpha_0/m_p v^2)^{1/4}$. As $\alpha_0 \simeq a^3$, $b_0 \gg a$ at small temperatures $T < 10^4 \, \rm K$. This approximation breaks down for electron scattering off neutral atoms.  As this case  requires $k b_0 \gg 1$, we do not satisfy this condition except at very high temperature.

It is clear that many  different approximations might be needed   to study this 
scattering  for a large range of temperatures.  Generically, the neutral atom is coupled
to charged particles at high redshifts and could decouple before recombination.  One of  our main aim is
to determine the redshift of decoupling and there are interesting observational
consequences when $z< 10^6$. For the redshifts of interest, Eq.~(\ref{eq:bornapp}) shows that the suitable 
approximation is $k a \ll 1$. This is the limit of low energy scattering 
which can be studied using partial wave analysis. In this limit, only
the $\ell =0$ partial wave contributes significantly\footnote{the scattering amplitude  for 
$\ell \ne  0$ scales as $(ka)^{2\ell}$} and the cross section is independent of 
the energy and the scattering angle. For scattering of a proton/electron off a neutral atom in its ground state,  the effective interaction potential  $U = e^2 \alpha/r^4$. The $s$-wave cross section of momentum exchange  of these atoms with electrons, in the center-of-mass frame,    is (see e.g. section~132 of \cite{1965qume.book.....L}):
\begin{eqnarray}
\sigma_{\rm hed} & \simeq & 4 \pi a_{\rm hed}^3/a_{\rm h} \\
\sigma_{\rm dd} & \simeq & 4 \pi a_{\rm dd}^3/a_{\rm h}
\label{eq:ddhe_intrate}
\end{eqnarray}
Here $a_{\rm hed} = \hbar^2/(Z^2 e^2 m_{\rm he})$ and $a_{\rm h} = \hbar^2/(e^2 m_{\rm e})$ ($Z=2$) and $a_{\rm dd} = \hbar^2/(e^2 m_{\rm dd})$. For scattering of these atoms  off protons, the mass of 
electron is replaced by the mass of proton in Eq.~(\ref{eq:ddhe_intrate}) and  
these cross sections are larger by a factor $(m_p/m_e)$ \footnote{to be more exact, $m_{\rm he}$  should be replaced by the reduced mass of the atom and $m_e$ or $m_p$ should be replaced by the reduced mass of the scattering particles. However, as the mass of the dark matter particle is much larger than the helium nucleus and the mass of the HeD atom is much larger than the  proton mass, this is a good approximation}. It should be noted
that the cross sections are independent of the relative velocity. In the literature, atomic dark matter  in the dark sector has been
considered  which might yield velocity-dependent cross section  (e.g. \cite{2013PhRvD..87j3515C}). 

\subsection{The rate of momentum loss} \label{sec:momloss}
For computing the time scales $\tau_{\rm db}$ and $\tau_{\rm bd}$ defined in section ~\ref{sec:evol_cou}, we need to compute the rate at which the momentum is
lost in the lab frame. This rate is defined as (e.g. \cite{1969mech.book.....L}):
\begin{equation}
\nu_1 = {v_1 \over p} {dp\over dl}
\end{equation}
Here $p$ is the momentum of the incident particle and $v_1$ its velocity in the lab frame. The target particle is assumed to be at rest in this frame so $v_1$ is the relative velocity between the two particles. $dp/dl$ the loss of momentum of the incident particle per unit length. It can be shown that:
\begin{equation}
  \nu_1 = {n_2 v_1 m_2 \over (m_1+m_2)}{d\sigma\over d\Omega}(\chi) (1-\cos\chi)
  \label{eq:scattrate}
\end{equation}
Here $n_2$ and $m_2$  are  the number density and  the mass of the target particle, respectively. $\chi$ is the angle of scattering and $d\sigma/d\Omega$ is the differential  cross section of scattering in the center of mass frame. Integrating Eq.~(\ref{eq:scattrate})  over the solid angle  and averaging
over the distribution of $v_1$ we obtain the final expression of the rate of momentum
loss. Similarly,  we can obtain $\nu_2$, the rate of the inverse process. It should be
noted that $\nu_1/\nu_2 = \rho_2/\rho_1$, where $\rho = n m$ is the mass density.
Using Eqs.~(\ref{eq:thetab}) and~(\ref{eq:thetac}), it can readily be shown that the rate of momentum exchange between the two particles per unit volume---$\rho_c\dot{\theta_c}$  and  $\rho_b\dot{\theta_b}$---is the same for the forward and the inverse process, as required by the conservation of momentum.

Eq.~(\ref{eq:scattrate}) allows us to  compute the time scales  $\tau_{\rm bd}$ and $\tau_{\rm db}$ which are inverse of the rate of momentum loss. Eq.~(\ref{eq:col_int})
gives $\tau_{\rm bd}$ for coulomb interaction.

For the other two models we discuss, the cross section is independent of velocity
and the angle of scattering (Eq.~(\ref{eq:ddhe_intrate})). For these cases, averaging over the thermal distribution of velocity, we get:
\begin{equation}
  \tau_{\rm db}^{-1} = {2 \over \pi^{1/2}} \left({n_{b} m  \over m + m_{\rm dm}}\right) \sigma \left [ {2k_b  T \over m_r} \right ]^{1/2}
  \label{eq:inttime}
\end{equation}
Here $n_{\rm dm}$ and $m_{\rm dm}$ correspond to number density and mass of
either  the  DD or HeD atom.  $m$ is the mass of either electron, proton, or Helium nucleus
and $\sigma$ is the relevant cross section.  $m_r$ is the reduced mass of the scattering particles. As discussed above, $\tau_{\rm bd}^{-1} = \tau_{\rm db}^{-1} \rho_{\rm dm}/\rho_b$. 

We note that the scattering cross section might not correspond to a single
process  but could be an effective cross section as the dark matter particle (which refers to either CCDM particle or dark matter atom) interacts electromagnetically  with free
electrons, protons and helium nuclei. The cross section of interaction between
the dark matter and these particles is not the same. The impact of all these interactions can be captured by defining   an effective time of interaction,  $\tau_{\rm db}^{-1} = \sum_i 1/\tau_{\rm di}$, where $i$ runs over these species,   because 
all these particles share the momentum exchange with the dark matter particle. This also
allows us to compute an effective scattering cross section $\sigma_{\rm db}$ in terms of the scattering cross sections with different charged particles. Similarly,
$\sigma_{\rm bd}$ can also be defined. In our case, the scattering is dominated by a single process for each case: CCDM-electron, DD-proton, and HeD-proton.

The dark matter and baryons get strongly coupled when their interaction time
scales become comparable to the expansion rate. The ratio of the  interaction rate and the expansion rate, $H$, is:
\begin{eqnarray}
 {1 \over \tau_{\rm db}^{\rm ccdm} H} & \simeq  & 0.1 \left ({5 \, {\rm MeV} \over m_{\rm ccdm}} \right ) \left ( {q_{\rm ccdm} \over 10^{-6} e} \right )^2 \left ({5\times 10^4 \over z} \right)^{1/2} \label{eq:ccdmhr}\\
    {1\over \tau_{\rm db}^{\rm dd} H} & \simeq & 6 \times 10^{-2}  \left ({25 \, {\rm GeV} \over m_{\rm dd}} \right )^4 \left( {z \over 5\times 10^4} \right)^{3/2} \label{eq:ddhr}\\
    {1\over \tau_{\rm db}^{\rm hed}H} & \simeq & 1.2 \times 10^{-3}  \left ({5 \, {\rm TeV} \over m_{\rm hed}} \right ) \left( {z \over 5\times 10^4} \right)^{3/2} \label{eq:hedhr}
\end{eqnarray}
Here we have used the expansion rate in the radiation-dominated era, $H(z) \propto (1+z)^2$. For the CCDM model, the ratio of  the interaction  and expansion rate increases with time and is a constant in the
matter dominated era. For the other two cases, the rate of interaction
is greater than the expansion rate at early times.

\section{Evolution of coupled dark matter-baryon  density and velocity  perturbations} \label{sec:evol_cou}
The interaction of all the proposed dark matter candidates occurs via momentum transfer with baryons. The dark matter-photon coupling is sub-dominant to the electron-photon coupling in all the cases we consider and therefore 
we neglect it. The main impact of the additional  interaction  is to alter  the evolution of density ($\delta$) and velocity  perturbations  ($\theta$) of  the coupled  photon-baryons fluid along with the interacting component of the  dark matter. The relevant equations  in the Newtonian/Conformal gauge  are (for details see e.g.  \cite{1995ApJ...455....7M}):  
\begin{eqnarray}
\dot{\delta_c}  & = &  - \theta_c + 3 \dot{\phi} \\
\dot{\theta_c} &  = &  - \frac{\dot{a}}{a} \theta_c + k^2 \psi + \frac{a}{\tau_{\rm db}}\left (\theta_b - \theta_c \right ) \label{eq:thetac}\\
\dot{\delta_b}  & = & -\theta_b + 3\dot{\phi}\\
\dot{\theta_b} &=& - \frac{\dot{a}}{a} \theta_b +k^2 \psi + \frac{aR}{\tau_{\rm e\gamma}}\left (\theta_\gamma - \theta_b \right ) + \frac{a}{\tau_{\rm bd}}\left (\theta_c - \theta_b \right )  \label{eq:thetab}  \\
\dot{\delta_\gamma}  & = & -{4\over 3}\theta_\gamma + 4\dot{\phi}\\
\dot{\theta_\gamma} &=& k^2\left({\delta_\gamma \over 4}-\sigma_\gamma\right ) + k^2 \psi + \frac{a}{\tau_{\rm e\gamma}}\left (\theta_b - \theta_\gamma \right )   \label{eq:thetagamma} 
\end{eqnarray}
Here the dot corresponds to the derivative with respect to the  conformal time, $d\eta = dt/a$. $\tau_{\rm e\gamma} = 1/(n_e \sigma_{\rm T} c)$ determines the interaction time
scale between baryons  and photons while $R = 4\rho_{\gamma}/(3\rho_b)$ denotes  the relative inertia of baryons and photons in the scattering process and this factor ensures: (a) the entire baryonic fluid shares the momentum
exchange between photons and electrons and (b) this scattering process is 
momentum conserving.   $\tau_{\rm bd}$ and $\tau_{\rm db}$ are  the interaction time scales  between baryons and dark matter. Here it is again implicitly assumed the momentum exchange between the two particles is shared by the entire baryonic fluid. These time scales are derived and discussed in section~\ref{sec:momloss}. As shown in section~\ref{sec:momloss},  $\tau_{\rm bd}^{-1} = \rho_{\rm dm}\tau_{\rm db}^{-1}/\rho_b$.  Here the mass density of interacting component of dark matter, $\rho_{\rm dm} = m_{\rm dm} n_{\rm dm}$, with $m_{\rm dm}$  and  $n_{\rm dm}$ being  the  mass and the number density interacting dark matter particles, respectively.   
In our study we assume all the species to be at the same temperature (see later
sections for details).   $\psi$ and  $\phi$ are cosmological potentials in the Newtonian/conformal  gauge,  $\sigma_\gamma$ is the second moment of the photon distribution,  and 
all the other variables have their usual definitions  (for details see \cite{1995ApJ...455....7M}).  The rest of the equations of 
the coupled multi-component fluid along with Einstein's equations remain the 
same. We solve these equations by modifying the publicly-available code CLASS \cite{Blas:2011rf}.

The main impact of the models considered here is captured by the last terms
in Eqs.~(\ref{eq:thetac}) and~(\ref{eq:thetab}). To solve these equations along with
Eq.~(\ref{eq:thetagamma}), we use the
tight coupling approximation whenever the relevant time scales are much  shorter than the
expansion  time scale; the details of this approximation are discussed in the next section.  When the time scales of interaction between
baryons and dark matter exceed the expansion time scale, Eq.~(\ref{eq:thetac})  can be integrated
directly and Eq.~(\ref{eq:thetab}) can be merged with Eq.~(\ref{eq:thetagamma})
in the usual  tight-coupling expansion in $\tau_{\rm e\gamma}$.

\subsection{Dynamics  of baryon-photon-dark matter fluid: tight-coupling approximation} \label{sec:tighcou}

We draw upon Eqs.~(\ref{eq:thetac})--(\ref{eq:thetagamma}) in this section. We first briefly discuss the usual case in which there is no coupling between dark matter and
baryons. In this case, the dynamics of
the coupled baryon-photon fluid is determined by a two time scales $\tau_{\rm e\gamma}$ and
$\tau_{\rm e\gamma}/R$; $R=4\rho_\gamma/(3\rho_b) \gg 1$ in the early universe as it scales as $1/a$. Both these time scales are much shorter than the expansion time scale
before the recombination sets in close to $z\simeq 1100$,  which makes it  difficult to numerically  solve the coupled evolution of the  baryon-photon fluid in the early universe. However, for the scales of interest, the photon-baryon fluid can be treated using tight-coupling approximation for $z \gg 1000$. In this
approximation, the two fluids oscillate with a common sound velocity: $c_s \simeq c\sqrt{R/(3(1+R))}$. To take into account both the common oscillation of the fluid and Silk damping owing to
the finite mean free path of the photon, one needs to solve the coupled photon-baryon equations up to  second order in $\tau_{\rm e\gamma}$.  The   exact equations can be numerically  solved  when  $\tau_{e\gamma} \simeq  H^{-1}$. This approach is adopted in all the CMB codes that numerically solve the
coupled baryon-photon evolution (e.g. \cite{1995ApJ...455....7M}).

The interaction between baryons and dark matter introduces additional complications. For two
of the models we consider, DD and HeD, the tight-coupling approximation also applies to
the coupled baryon-dark matter system in the early universe. In the early universe, 
all the four time scales---$\tau_{e\gamma}$, $\tau_{e\gamma}/R$, $\tau_{\rm db}$ and $\tau_{\rm bd}$---are shorter than the expansion time scales, and the baryon-photon-dark matter fluids can be  treated
as tightly coupled.

In this case, $\theta_b-\theta_\gamma$ in Eq.~(\ref{eq:thetagamma}) and $\theta_b - \theta_c$
in Eq.~(\ref{eq:thetac}) can be expanded in $\tau_{\rm e\gamma}$ and $\tau_{\rm db}$, respectively,
and substituted in Eq.~(\ref{eq:thetab}). This gives us:
\begin{equation}
  (1+R+R')\dot\theta_b+{\dot a \over a}(\theta_b +R'\theta_c)-k^2R\left ({\delta_\gamma \over 4}-\sigma_\gamma \right )+R(\dot\theta_\gamma - \dot\theta_b)+R'(\dot\theta_c - \dot\theta_b) = (1+R+R')k^2\psi
  \label{eq:tightcoup}
\end{equation}
Here $R' = \tau_{\rm db}/\tau_{\rm bd} = \rho_{\rm dm}/\rho_b$. Eq.~(\ref{eq:tightcoup}) is exact. To obtain the equation which is correct to first order in
$\tau_{e\gamma}$ and $\tau_{\rm db}$, we put $\sigma_\gamma = 0$ and $\theta_b = \theta_\gamma=\theta_c$. In the absence of dark matter-baryon coupling  $\dot\theta_\gamma - \dot\theta_b$  in Eq.~(\ref{eq:tightcoup}) needs
to be  expanded to first order in $\tau_{e\gamma}$  to correctly  account for photon diffusion damping (Silk damping). We note that this also suffices in our case  as the diffusion damping owing to
baryon-dark matter coupling impacts much  smaller scales \footnote{The typical scales
  impacted by Silk damping owing to photon diffusion, $\lambda_{\rm b\gamma}(t)  \simeq c \sqrt{\tau_{\rm e\gamma} H}$  (see e.g. \cite{2003moco.book.....D}) while the corresponding scale for diffusion owing to dark matter-baryon coupling is
  $\lambda_{\rm bd}(t)  \simeq v \sqrt{\tau_{\rm e\gamma} H}$. For all the cases we discuss here
  $\lambda_{\rm b\gamma}  \gg \lambda_{\rm bd}$.}; the evolution of $\theta_\gamma$ can  be obtained after eliminating $\theta_\gamma - \theta_b$ in Eqs.~(\ref{eq:thetagamma}) and~(\ref{eq:thetab}) (e.g. Equations~70, 74, and~75 of \cite{1995ApJ...455....7M}). Similarly, the suitable equation for the evolution
of $\theta_c$ can be obtained from Eqs.~(\ref{eq:thetac}) and~(\ref{eq:thetab}). Eq.~(\ref{eq:tightcoup}) allows
us to compute the approximate sound speed in the coupled baryon-photon-dark matter fluid: $c_s^2  \simeq R/(3(1+R+R'))$. 

The DD and HeD models we consider  in this paper correspond to cases in which the decoupling of dark matter and baryons occurs in the redshift range  $z > 10^4\hbox{--}10^6$.  If all the cold dark matter is in the form of these atoms, which is allowed (section~\ref{sec:Compare}),  $R' = \rho_c/\rho_b \simeq 5.5$. For comparison, $R\simeq 100$ at $z\simeq 10^5$. This means  $R' \ll R$ for a majority of the cases we consider.

In the  CCDM model, the baryon-dark matter coupling is weak at early  times. This allows
us to solve   the dynamics of dark matter and baryons-photon fluid separately   at early time.  When this coupling
becomes large (at $z \le 5000$ for the models we study), the system of equations   can be solved in the tight-coupling approximation. For all the models that are compatible with cosmological observables, the fraction
of CCDM dark matter, $f_{\rm ccdm}$, is a few  percent. This implies $R' \ll R$ for this
case also.

For $R' \ll R$ and $R' \ll 1$, it follows from Eq.~(\ref{eq:tightcoup}) that  the  dark matter-baryon coupling has negligible impact on the dynamics of the photon-baryons fluid while the dynamics of the interacting component of the dark matter is significantly
affected by the coupling.

\subsection{Matter power spectrum and CMB anisotropies}

The tight coupling approximation can be used to gauge the impact of the additional coupling on
the matter power spectrum and CMB anisotropies.

In the usual $\Lambda$CDM case,  perturbations (in the Newtonian gauge)  in all the matter components are constant at superhorizon scales. At sub-horizon scales, the cold dark matter perturbations grow logarithmically
in the radiation-dominated era and as $\eta^2$ in the matter-dominated era. The photons and baryons are tightly
coupled before recombination era and their coupled perturbations at sub-horizon scales  oscillate with a constant amplitude (e.g. \cite{2003moco.book.....D}). 

\begin{figure}
        \centering
        \begin{minipage}{0.49\textwidth}
                \centering
                \includegraphics[width=1.0\linewidth]{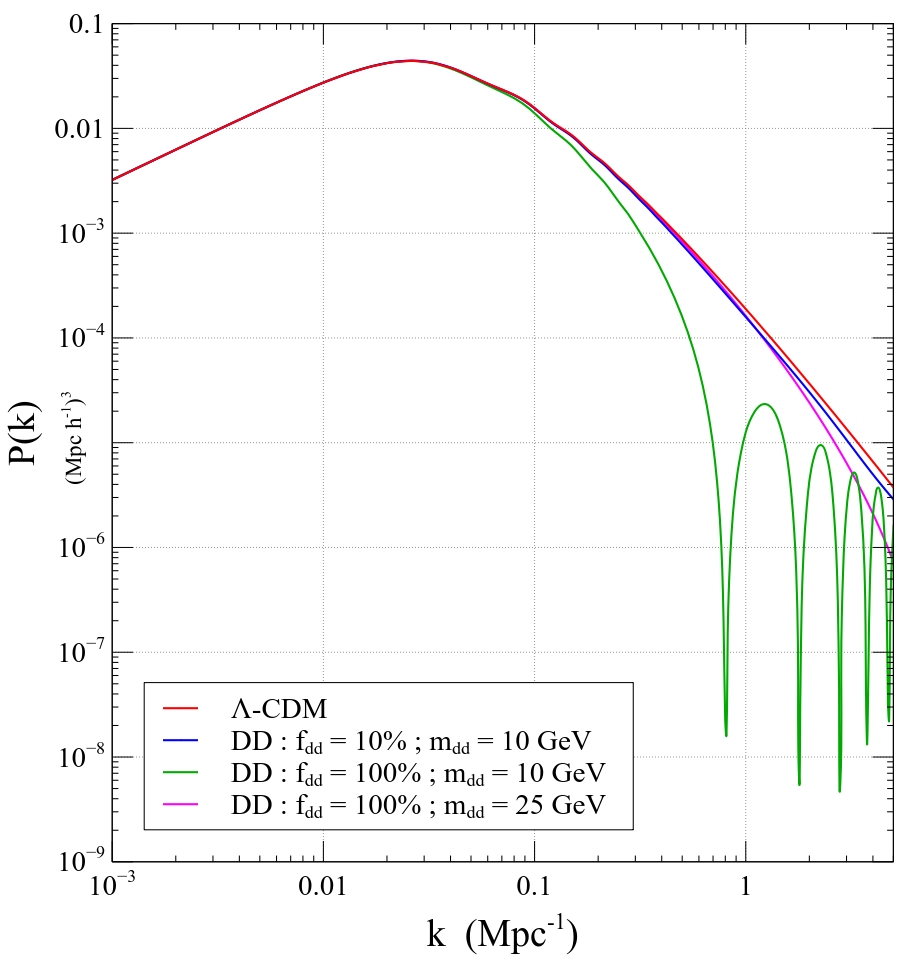}
        \end{minipage}\hfill
        \centering
        \begin{minipage}{0.49\textwidth}
                \centering
                \includegraphics[width=1.0\linewidth]{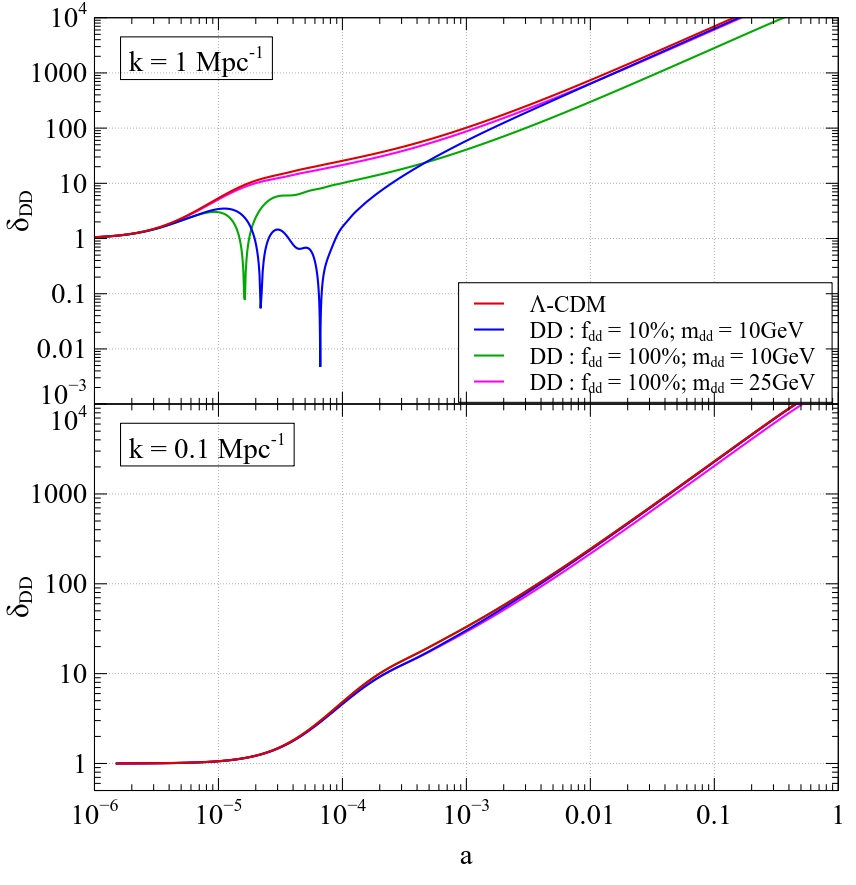}
        \end{minipage}\hfill
\caption{Left panel: The matter power spectra $P(k)$  ($z=1000$) are shown for many  parameters of the $\rm DD$ model, along with the results of the usual $\Lambda$CDM model. Right Panel: The time evolution of $\delta_{\rm DD}$ is displayed for  two Fourier modes,  $k=0.1 \, \rm Mpc^{-1}$ and $k=1 \, \rm Mpc^{-1}$.
}
\label{fig:dd_pow}
\end{figure}

\begin{figure}
        \centering
        \begin{minipage}{0.49\textwidth}
                \centering
                \includegraphics[width=1.0\linewidth]{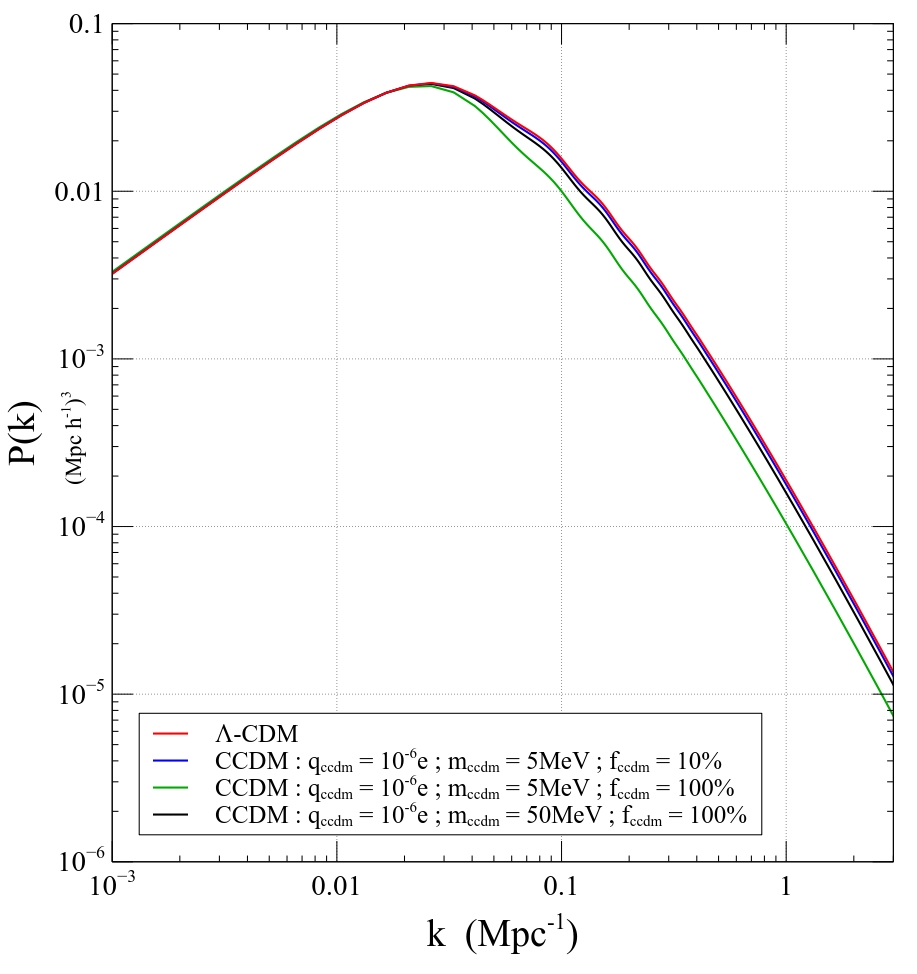}
        \end{minipage}\hfill
        \centering
        \begin{minipage}{0.49\textwidth}
                \centering
                \includegraphics[width=1.0\linewidth]{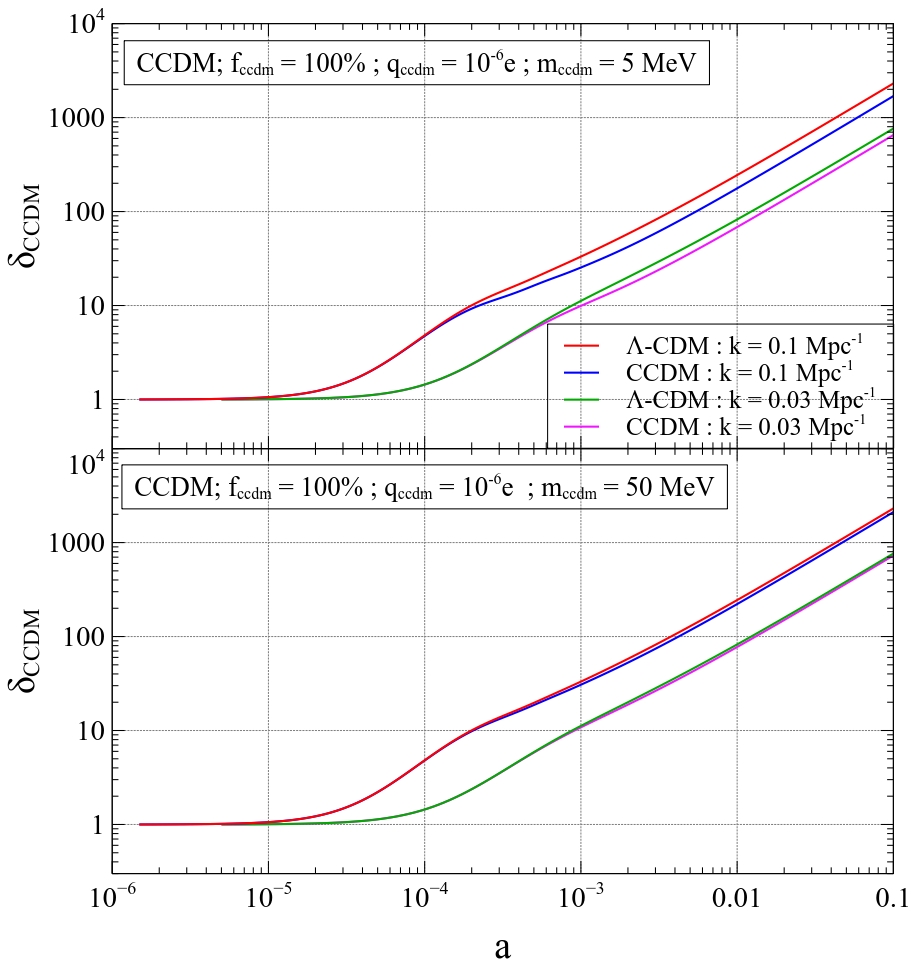}
        \end{minipage}\hfill
\caption{Left panel: The matter power spectra $P(k)$  are shown for several  CCDM models at $z=1000$.  The  $\Lambda$CDM matter power spectrum  is also shown for comparison. Right Panel: The time evolution of density 
contrast of the interacting component of the dark matter $\delta_{\rm CCDM}$ is 
displayed for two Fourier 
modes,  $k=0.1 \, \rm Mpc^{-1}$ and $k=0.03 \, \rm Mpc^{-1}$.
}
\label{fig:ccdm_pow}
\end{figure}

In the presence of additional  dark matter-baryon coupling, a fraction (or all) of dark  matter could behave as
baryon-photon fluid. This means that  the perturbations in this dark  matter component cannot grow either
logarithmically or as $\eta^2$ before the era of recombination. This causes a suppression in the matter power spectrum at scales that
are sub-horizon when the coupling is strong. For the DD and HeD models, the coupling is strong at only early times
and therefore the smaller scales are affected. For the CCDM  case, the coupling is strong at latter times which impacts scales that enter the horizon around the time of recombination. This is clearly seen in the Figures~\ref{fig:dd_pow} and~\ref{fig:ccdm_pow}.


The CMB anisotropies are sensitive to perturbations close to the epoch of recombination, $\eta \simeq \eta_r$.
The angular scale of  the observed anisotropies $\ell$  correspond approximately to  the Fourier mode  of perturbations $k$ as $\ell \simeq k \eta_0$, where $\eta_0 \simeq  14400 \, \rm Mpc$ correspond to the conformal time at the present. As Planck can measure anisotropies for angular scales $\ell < 2500$, the CMB anisotropies carry information of  Fourier modes, $k < 0.2 \, \rm Mpc^{-1}$. The primary source term for the CMB anisotropies is (e.g. \cite{2005pfc..book.....M,2004IJTP...43..623M}):
\begin{equation}
  {\cal S}(\eta_r,k) \simeq \phi(\eta_r,k) + {\delta_\gamma(\eta_r,k) \over 4}
\end{equation}
At large scales, $k\eta_r \ll 1$, the source term , ${\cal S}(\eta_r,k) =  \phi(\eta_r,k)/3$ yields the
well-known Sachs-Wolfe effect. These scales are not affected by the  baryon-dark matter coupling. The source
term at smaller scales, $k\eta_r > 1$, is difficult to compute analytically. Approximate expressions
have been used in the literature to capture the essential physics at these scales. We use the analysis presented by Mukhanov \cite{2004IJTP...43..623M} to underline the impact of the additional coupling on CMB anisotropies. Following \cite{2004IJTP...43..623M}, the source term for $k\eta_r >$  is:
\begin{equation}
  {\cal S}(\eta_r,k) \simeq \left [ {\log(b_1 k\eta_{\rm eq}) \over (b_2 k\eta_{\rm eq})^2}\left(1-{1 \over 3c_s^2}\right) + 2\sqrt{c_s} \cos\left(k\int_0^{\eta_r} c_s d\eta\right) \exp(-k/k_D)^2 \right ] \phi(\eta_i,k)
  \label{eq:cmbani}
\end{equation}
Here $\eta_i$ corresponds to an initial time at which $k\eta_i \ll 1$. $\eta_{\rm eq}$ is the conformal time
at matter-radiation equality and $c_s \simeq \sqrt{R/(3+3R)}$ is the sound velocity of the coupled baryon-photon fluid at the recombination era. $k_D^2 \simeq 1/(\eta \tau_{e\gamma})$ gives the Silk damping scale. $b_1$ and $b_2$ are constants whose values depend on $k$.   Eq.~(\ref{eq:cmbani}) allows us to assess the impact of baryon-photon coupling on the observed CMB anisotropies. First, as already discussed above, the Silk damping scale $k_D$ is not affected by this coupling.

In DD and HeD models, the decoupling occurs at early times and therefore  quantities such as $c_s$ whose value only close to the recombination era
contribute significantly to CMB anisotropies are not affected. The main impact comes from the first
term on the RHS of Eq.~(\ref{eq:cmbani}) which arises from the time evolution of the gravitational potential $\phi(\eta,k)$. $\phi(\eta,k)$ is constant at superhorizon scales in the radiation-dominated era and is constant at all scales in
the matter-dominated era. Neglecting the impact of cold dark matter perturbations, the potential
decays as $1/\eta^2$ for scales that enter the horizon during the radiation-dominated era. The perturbations of the cold dark matter grow logarithmically for sub-horizon scales
during the radiation-dominated era. And when this effect is taken into account, the potential
falls slower than $1/\eta^2$; this is the origin of the  log term in Eq.~(\ref{eq:cmbani}). For DD and HeD
models, the cold dark matter perturbation cannot grow when the baryon-dark matter coupling is strong, which
diminishes the value of log term in Eq.~(\ref{eq:cmbani}). As the values of $b_1$ and $b_2$ are  strong functions  of $k$ (e.g. \cite{2005pfc..book.....M,2004IJTP...43..623M,2003moco.book.....D}), we do not
try to estimate it analytically here but only present numerical results for
the matter power spectrum (Figures~\ref{fig:dd_pow} and~\ref{fig:ccdm_pow}).

In the CCDM case, the baryon-dark matter coupling becomes stronger at later time and its main impact occurs close to the time of decoupling. CMB data strongly constrains the angle subtended on by the 
sound horizon at the epoch of recombination, $\theta_\star = r_s/D_A$, where $r_s \simeq \eta_r \sqrt{R/(3(1+R))}$ and $D_A$ is the angular diameter distance to the epoch of recombination.  Planck
data yields  $\theta_\star = 0.59643\pm0.00026$ (in degrees) \cite{Planck2018}.   In the usual case, $R=4\rho_\gamma/(3\rho_b)$ and $R\simeq 1.2$
at $z\simeq 1000$.  If all the dark matter is in the form of CCDM particles, and these particles
are strongly coupled to baryons at recombination, $r_s \simeq  \eta \sqrt{R/(3(1+R+R'))}$, which is radically
different from the usual value as $R' \simeq 6$.  This is readily ruled out by the CMB data. Therefore, we
only consider models for which  $R' \ll 1$, or the interacting dark matter must be a small fraction of baryons.  In such cases,  the term corresponding the potential evolution in Eq.~(\ref{eq:cmbani})
is minimally affected and the main impact of CCDM models on CMB anisotropies is owing to
the change in $c_s$ in the second term on the RHS  because of  non-zero $R'$.

\subsubsection{Numerical implimentation in CLASS}

It follows  from the discussion above that for the DD/HeD cases, the dark matter
is tightly coupled at high redshifts and then decouples when the interaction
rate between the dark matter and baryons falls below the expansion rate. On the other hand, the coupling becomes stronger at later time in CCDM case.

For numerical stability, we switch from  the tight-coupling approximation to  the exact equations when the dominant interaction rate is  100 times the expansion rate for the CCDM model.   As noted above, the CCDM and baryon cannot be tightly coupled  at the last scattering surface for  viable cosmological models, if the fraction of CCDM is substantial. This  allows us  to   use the exact equations  for a majority of the cases we study.   For the DD/HeD cases we make the switch  when the interaction rate
is  20 times the expansion rate. This allows us to evolve wavenumbers  $k \lesssim 10 \, \rm Mpc^{-1}$ without glitches. We note that this prescription is similar to and in addition to the
usual case in which the photon-baryon plasma makes transition from tightly
coupled to photon decoupling close to the epoch of recombination.

The initial conditions for the CCDM case are the same as in the usual case
because the milli-charged dark matter particle is upcoupled from the
baryons at early times. In the other two cases, a fraction of the dark matter is tightly
coupled to baryons at early times so the initial condition for bulk velocity of this component  is derived
from the tight-coupling relation: $\theta_c = \theta_b$. The other initial
conditions remain the same.

In section~\ref{sec:appendix} we discuss the formation of DD and HeD atoms. 
The formation redshift of these atoms is  before the initial conditions
are set in CLASS: radiation domination era with scales of interest ($k \lesssim 10 \, \rm Mpc^{-1}$) well
outside the horizon. However, we note that the initial conditions are immune to
whether the dark matter is in free (ionized) form or in atomic form because
the interaction rate of both these components far exceeds the expansion rate
when initial conditions are set. All the components that are coupled to the
baryon-photon plasma share a common bulk velocity with it, as  follows
from the discussion on tight-coupling approximation in the foregoing. This
situation is akin to how different stages of ionization of helium nuclei are
incorporated in numerical codes in the usual $\Lambda$CDM case---the neutral helium
is treated the same way as ionized helium because the interaction rates of both
components with electrons/protons far exceeds the expansion rate.

\section{Matter power spectrum}
\label{sec:Compare}

The last section provides a general framework to study cosmological
implications of additional baryon-dark matter interaction. 
In this section we focus on  specific models and  their matter power spectra.  We build on the discussion in  Section~\ref{sec:tighcou} and provide
numerical results. 

It follows from Eqs.~(\ref{eq:thetac}) and~(\ref{eq:thetab}) that all the models tend towards the standard $\Lambda$CDM case when interaction rates fall  below the  expansion rates  at redshifts of interest: $\tau_{ij} \gg H^{-1}(z)$. The redshifts of interest can be determined from the following argument. When a scale is outside the horizon,  $k\eta < 1$, both the baryonic and dark matter
perturbations evolve in  the same way (for details e.g. \cite{1995ApJ...455....7M}).  If the baryons and dark matter are coupled during this phase but get decoupled before the scale enters the horizon, 
the impact of the additional coupling is negligible at that scale, i.e. large scales $k < 0.02 \, \rm Mpc^{-1}$ that enter the horizon after the epoch of recombination fall in this category. CMB anisotropies  and galaxy clustering  data are sensitive to scales $k < 0.2 \, \rm Mpc^{-1}$. If these scales enter the horizon when the baryon-dark matter coupling is strong then we expect observable
signatures  on the CMBR anisotropies and the  matter power spectrum. 
The redshifts of interest from
the point of view of observable signatures are $10^3 <  z < 10^5$, which corresponds to the time between the epoch of recombination and roughly the epoch
at which the mode $k \simeq 0.3 \, \rm Mpc^{-1}$ enters the horizon. 

{\it CCDM model}:   Eqs.~(\ref{eq:col_int}) and~(\ref{eq:ccdmhr}) show that the CCDM particles and baryons get coupled at late times.  To have a significant impact on 
cosmological observables this coupling must be strong  before the recombination era, which occurs when  either of the two time scales,  $\tau_{\rm db}$  and $\tau_{\rm bd}$, become comparable  to   the expansion time scale. 
In the two panels of Figure~\ref{fig:ccdm_pow},
we show the matter power spectra of the CCDM model and display the time evolution
of density perturbations of the interacting component of the dark matter, $\delta_{\rm CCDM}$,  for  a few Fourier modes  to glean  the important  features introduced by  CCDM models.

It follows from the discussion in section~\ref{sec:tighcou}  that the fraction of dark matter in the form of  CCDM, $f_{\rm CCDM} \ll 1$, if the  baryons and CCDM particles are tightly coupled  at $z\simeq 1000$. In Figure~\ref{fig:ccdm_pow}, we show a few cases
where the coupling is not strong enough to cause   coupled oscillation of the baryon-photon-dark matter fluid but strong  enough to significantly  alter the matter power spectrum.

As argued in section~\ref{sec:tighcou}, we expect the matter power to diminish in the presence of CCDM  as the baryon-dark matter
coupling prevents the growth of CDM perturbations. As the coupling becomes stronger with time, all the
scales that enter the horizon before the recombination era are affected in this case. We notice these features in
Figure~\ref{fig:ccdm_pow}. The main impact on CMB anisotropies in this case occurs owing to the change
the sound velocity of the coupled baryon-photon fluid at $z\simeq 1000$, as discussed in section~\ref{sec:tighcou}.

{\it DD atom}: The physics  of the  formation of this atom  and its interaction with baryons are 
discussed in detail in section~\ref{sec:appendix}. Eq.~(\ref{eq:ddhr}) shows that this atom is coupled
to baryons at early times and the coupling becomes weaker with time.  The scattering  cross section  of DD atoms off baryons  
scales as $1/m_{\rm dd}^3$ (Eq.~(\ref{eq:ddhe_intrate})) and therefore the redshift at which the decoupling occurs  is a very sensitive function of the dark matter mass. In Figure~\ref{fig:dd_pow} we 
show the matter power spectrum for several  models in which the DD atom
could constitute either a part or all of the cold dark matter.   
In the right panel of Figure~\ref{fig:dd_pow}, we also show the time evolution
of $\delta_{\rm DD}$, the density contrast of the interacting component of the 
dark matter.

The scales that are most  affected by dark matter-baryon interaction are those that are inside
the horizon when this coupling is still strong. For instance, if the decoupling occurs at $z\simeq 10^5$,  perturbations at  scales $k > 0.3 \rm Mpc^{-1}$ are   affected  by  this coupling.

From the discussion in section~\ref{sec:tighcou}, we can assess the impact of this additional coupling on the
evolution of different components in the tight-coupling approximation. The DD case allows for all the cold dark matter to be in the form of interacting
dark matter or  $R' = \rho_{\rm dd}/\rho_b \simeq 5.5$. For $R \gg R'$ and $R' \ll 1$,  the coupled
baryon-photon fluid is  negligibly affected by this new coupling  but perturbations of the
interacting component of the dark matter are driven by  the oscillations of the baryon-photon fluid. When
$R' \gtrsim  1$, the baryon-photon  perturbations are also significantly affected by this coupling. If the coupling is strong,  the three fluids oscillate together with a common
sound velocity, $c_s \simeq \sqrt{(R/(3(1+R+R'))}$, which  differs markedly from the sound velocity of the baryon-photon fluid. 

This discussion allows us to understand the features seen in Figure~\ref{fig:dd_pow}: (a) the matter power spectrum approaches the $\Lambda$CDM models as $m_{\rm dd}$ is increased and the
$f_{\rm DD}$ is lowered, (b) the interacting part of the dark matter behave as  the baryon-photon
fluid for scales that enter the horizon when the coupling is strong. This
explains the oscillations seen in the matter power spectrum for the model: $m_{\rm dd} = 10 \, \rm Gev$ and $f_{\rm DD} = 1$. This behaviour is also
evident in the time evolution of density perturbations (right panel), e.g.  
for $k = 1 \, \rm Mpc^{-1}$. For the models displayed in Figure~\ref{fig:dd_pow}, the decoupling occurs for  $z > 10^5$.

{\it HeD atom}:  Like the DD case, the coupling between baryons and HeD atoms is stronger  at high redshifts (Eq.~(\ref{eq:hedhr})). Therefore,
the physics of this model is similar to the DD case with a few notable differences. Unlike the DD case,  the dark matter-baryon scattering cross
section   does not depend on the mass of the dark matter particle (Eq.~(\ref{eq:ddhe_intrate})). The only
condition to ensure the formation and stability of this atom is  that the  dark matter  particle be  much heavier than the  helium nucleus. 
Therefore, unlike  CCDM and DD models, it is harder to tune the parameters of this model to seek 
agreement with observations and  this model is a more robust  representative 
of a paradigm that  admits electromagnetic  interaction between dark matter 
and baryons.

\begin{figure}
        \centering
        \begin{minipage}{0.49\textwidth}
                \centering
                \includegraphics[width=1.0\linewidth]{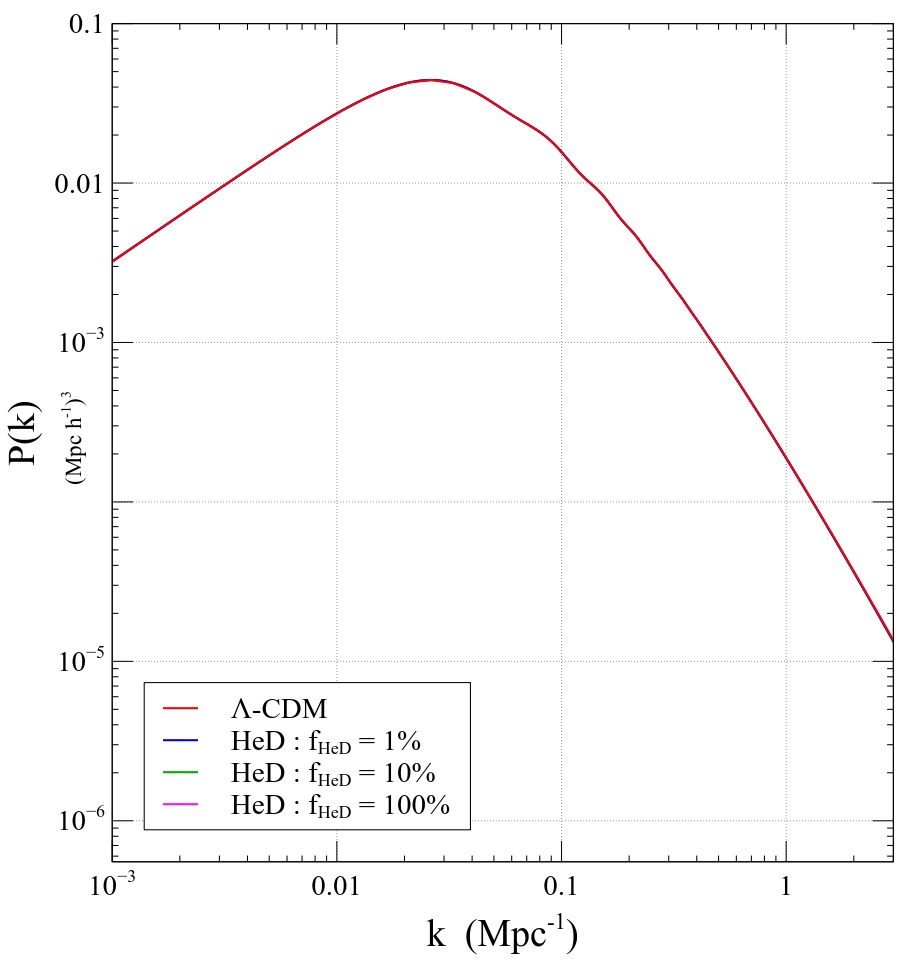}
        \end{minipage}\hfill
        \centering
        \begin{minipage}{0.49\textwidth}
                \centering
                \includegraphics[width=1.0\linewidth]{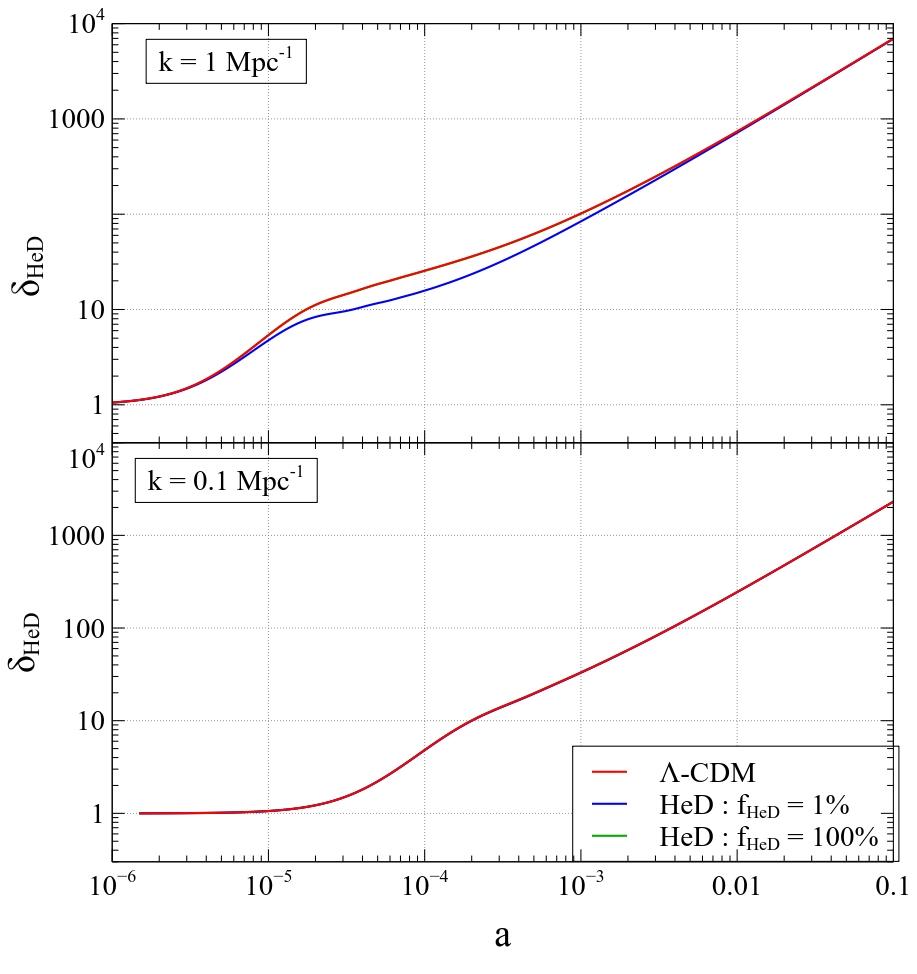}
        \end{minipage}\hfill
\caption{Left panel: The matter power spectra $P(k)$  are shown for $\rm HeD$ models for different values of $f_{\rm HeD}$ at $z=1000$, along with the  usual $\Lambda$CDM model. Right Panel: The time evolution of $\delta_{\rm HeD}$ is displayed for two Fourier
modes,  $k=0.1 \, \rm Mpc^{-1}$ and $k=1\, \rm Mpc^{-1}$.
}
\label{fig:hed_pow}
\end{figure}

An additional requirement in this model is that only a small fraction of $\rm ^4He$ be captured by dark matter
particles to form HeD atoms. From the current astronomical and CMB  data the primordial abundance of  $\rm ^4He$ could be  constrained to
better than one percent (e.g. \cite{Planck2018,2014MNRAS.445..778I} and references therein\footnote{CMB anisotropies  have a bearing  on the  Helium abundance by constraining the number of electrons captured by  Helium nuclei close to the epoch of recombination. In the HeD model, the charge neutrality of the universe requires us to have fewer electrons as compared to the usual model. The CMB anisotropies are sensitive to this deficit.}). Therefore, we  consider
models in which only $1\%$ of helium nuclei form atoms with dark matter.  This gives us: $n_{\rm hed}  \simeq 0.0008 n_b$ and  $m_{\rm dm} \simeq m_p  f_{\rm HeD} \rho_{\rm dm}n_b/(\rho_b n_{\rm hed})$, where $f_{\rm HeD}$ is the
fraction of dark matter in the form of HeD atoms. Using the Planck best-fit values of $\rho_{\rm dm}$ and $\rho_b$, we get, $m_{\rm hed} \simeq  7000 f_{\rm HeD} m_p$. Owing to the large mass of the particle that recombines
with helium nucleus to form the HeD atom, the interaction rate between baryons and HeD atoms is much smaller (Eq.~(\ref{eq:inttime})). For all the models we consider for different values of $f_{\rm HeD}$,
the matter power spectrum  is an excellent agreement with the $\Lambda$CDM model for $k \le 1\,\rm Mpc^{-1}$. In Figure~\ref{fig:hed_pow} we show the matter power spectra  and the  time evolution of two Fourier modes for a few cases.

\subsection{CMB anisotropies}

It follows from the discussion in section~\ref{sec:tighcou} that for the  DD and
HeD cases the impact of the additional baryon-dark matter coupling can be
captured by the change in the  matter power spectrum at scales that are inside the horizon during strong baryon-dark matter coupling. For the milli-charged particle, the change
in the velocity of sound close to the recombination provides an additional
discriminator. 

\begin{figure}
        \centering
        \begin{minipage}{0.49\textwidth}
                \centering
                \includegraphics[width=1.0\linewidth]{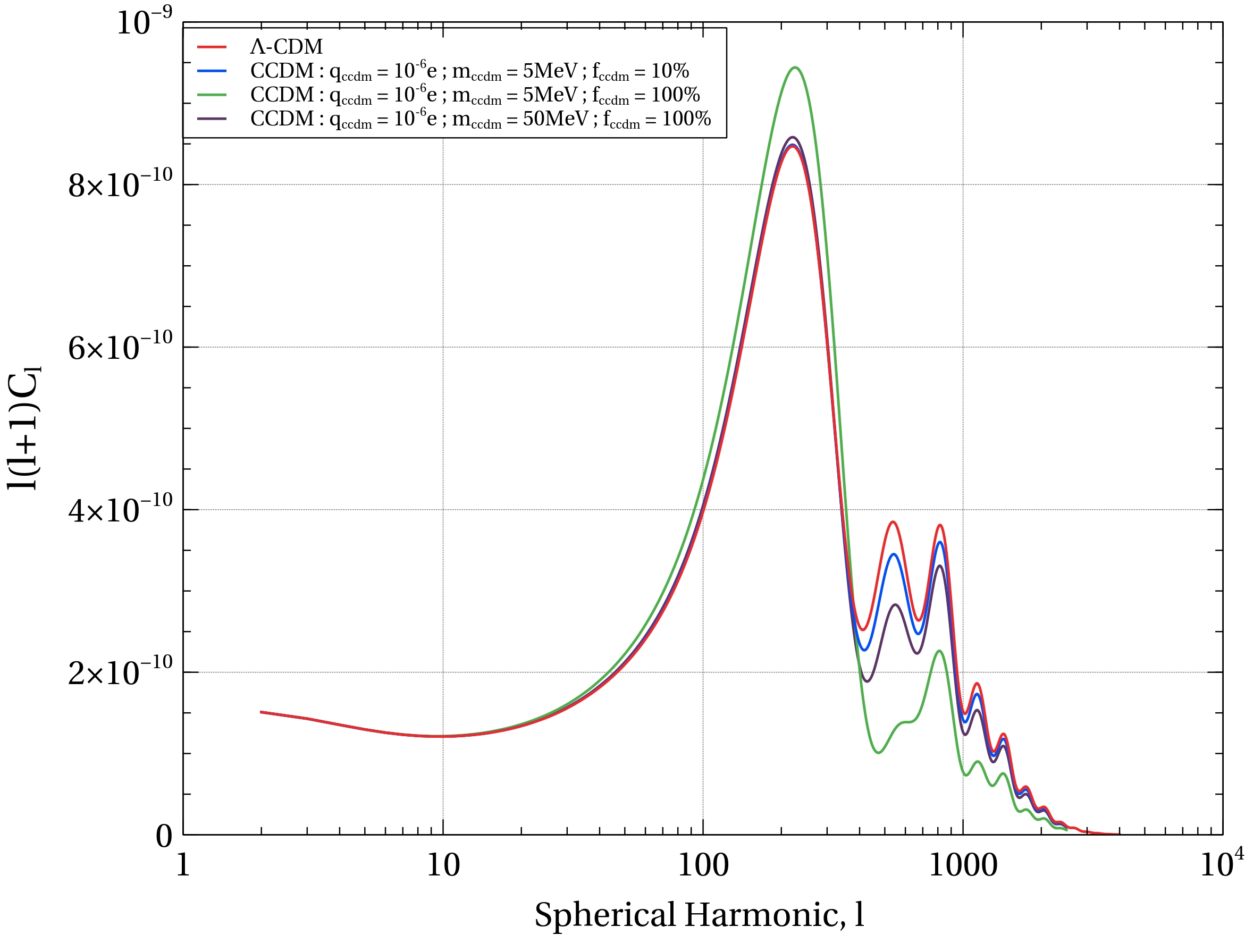}
        \end{minipage}\hfill
        \centering
        \begin{minipage}{0.49\textwidth}
                \centering
                \includegraphics[width=1.0\linewidth]{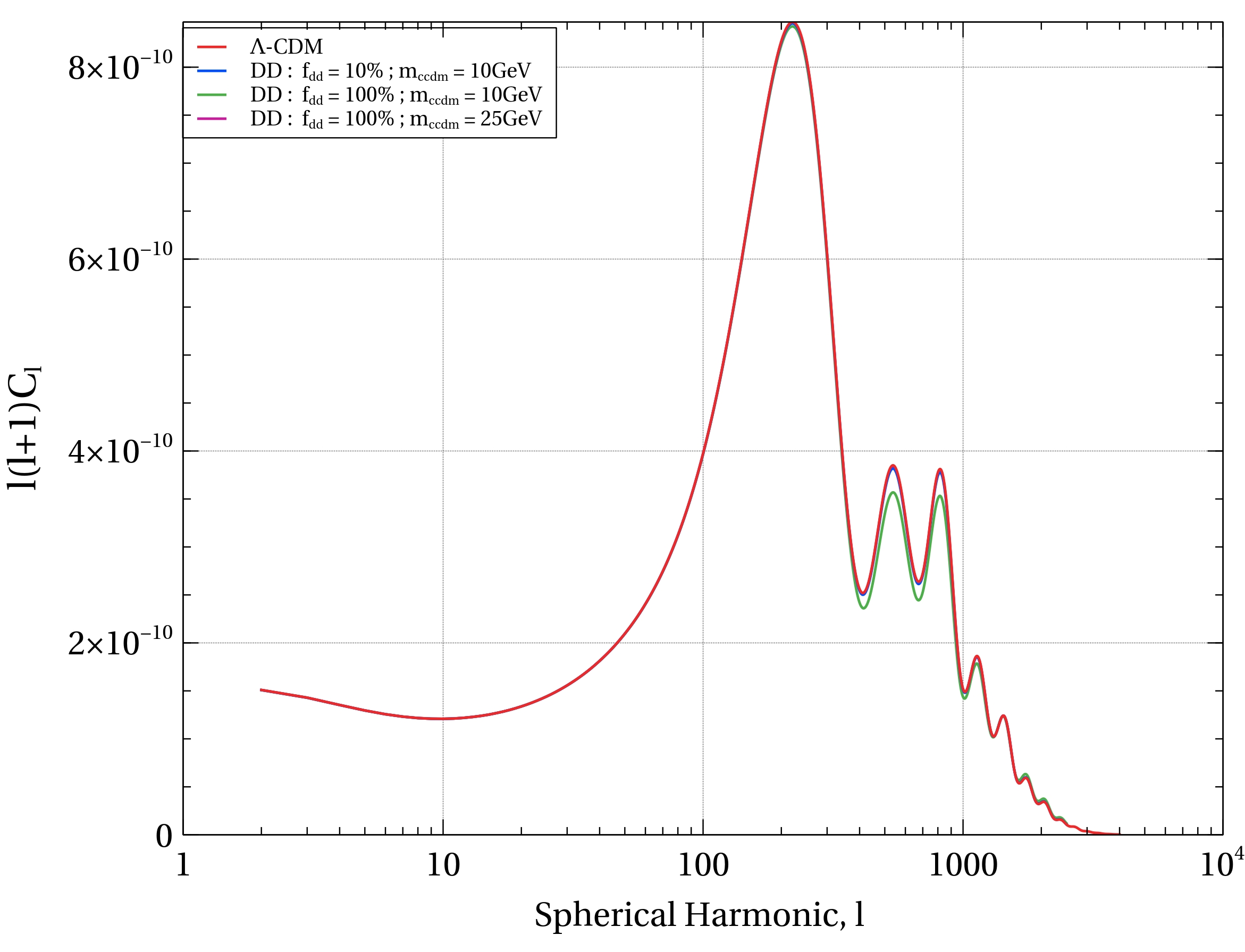}
        \end{minipage}\hfill
        \caption{The angular power spectra of CMB temperature anisotropy are shown for a few CCDM (DD) models  along with the   $\Lambda$CDM model.
}
\label{fig:cmb}
\end{figure}

The  angular scale
of CMB anisotropies ($l$)  can be related to scale of matter perturbations ($k$)  at last scattering surface  using  the approximation: $l \simeq k c \eta_0$, where   $\eta_0 =  \int dt/a(t)$ is the conformal time  at the present with $c \eta_0 \simeq 14000 \, \rm Mpc$. 
In Figure~\ref{fig:cmb}, we show the CMB temperature angular power
spectrum for a subset of models. As anticipated in the previous section,
the main impact of DD models is on large $l$  while the CCDM also  affects large angular scales $l \simeq 200$, which corresponds to wavenumbers   that  enter the horizon close to the
epoch of recombination. We do not show the expected CMB angular power
spectrum for the HeD model as, for permissible range of parameters, it is
indistinguishable from the usual model, in line with discussion in the previous section.

\section{Comparison with data}
\label{MCMC}
Our theoretical predictions can be compared to the existing CMB and galaxy
clustering data. The current Planck data is sensitive to $k < 0.2 \, \rm Mpc^{-1}$ \cite{Ade:2013zuv} and  the SDSS galaxy clustering data can also  be compared
to the prediction of linear theory for $k < 0.1 \, \rm Mpc^{-1}$ (e.g. \cite{Beutler:2016ixs}). As these scales are comparable and the CMB data is less affected by the
effects of non-linear clustering unlike the low-redshift SDSS data, we choose
to compare our results against Planck 2018 CMB  anisotropy data \cite{Planck2018}\footnote{For details of the data products and likelihoods see  https://wiki.cosmos.esa.int/planck-legacy-archive/index.php/CMB\_spectrum\_\%26\_Likelihood\_Code\#2018\_Likelihood}. 
The following Planck  likelihoods were used:  high-$l$ temperature and polarization anisotropies ($l = 30\hbox{--}2508$ for TT data and $l = 30\hbox{--}1996$ for TE and EE data), low-$l$ temperature and polarization data ($l = 2\hbox{--}29$), and CMB lensing reconstruction. 
The models were analysed using the MCMC code  MontePython \cite{2018ascl.soft05027B}.  

{\it Parameterization of models}: (a) {\it CCDM}: The three free parameters
of this model are: the charge of the dark matter particle, $q_{\rm ccdm}$, the mass of the dark matter particle, $m_{\rm ccdm}$, and the fraction of charged CDM, $f_{\rm ccdm}$.  As discussed in section~\ref{sec:tighcou}, if  CCDM-baryon coupling
becomes strong before the epoch of recombination, $f_{\rm ccdm} \ll 1$.
  Here we explore  models for  which the coupling is not strong enough
  to cause coupled CCDM-baryon oscillations. It is important to underline 
  that CMB anisotropies are extremely sensitive to behaviour of
  matter perturbations close to the epoch of recombination, (b) {\it DD}: In this case, there are two free parameters, the mass of the dark matter
particle,  $m_{\rm dd}$, and the fraction of DD atoms, $f_{\rm dd}$. This model
approaches the $\Lambda$CDM model when $m_{\rm dd} \rightarrow \infty$ and/or $f_{\rm dd} \rightarrow 0$. For MCMC analysis, we fix the value of $m_{\rm dd} = 10 \, \rm GeV$, which is close to the minimum mass allowed by charge-exchange stability (c) we do not carry out an MCMC analysis for the HeD case because, as noted above, all the models that are allowed by other  astrophysical data  are in excellent agreement with the $\Lambda$CDM model for
$k < 1 \, \rm Mpc^{-1}$ and therefore are not sensitive to the  CMB data.

\begin{figure}
\centering
\includegraphics[width=\textwidth]{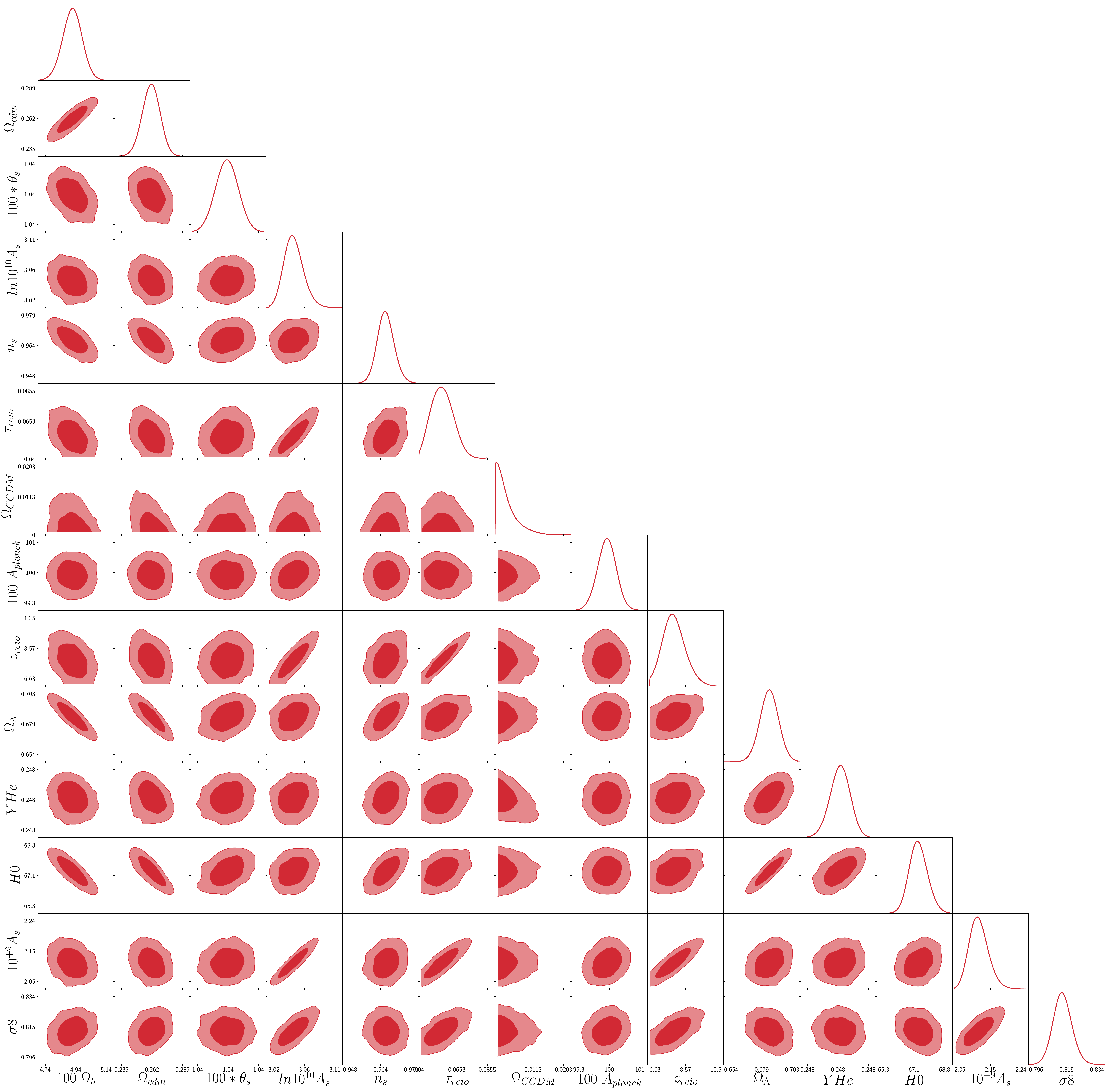}
\caption{Triangle graph from the MCMC analysis for the CCDM model is shown for $m_{\rm dm}=50 \,  \rm MeV$ and $q_{\rm ccdm} = 10^{-6}$e.}
\label{fig:ccdm_post}
\end{figure}

\begin{figure}
\centering
\includegraphics[width=\textwidth]{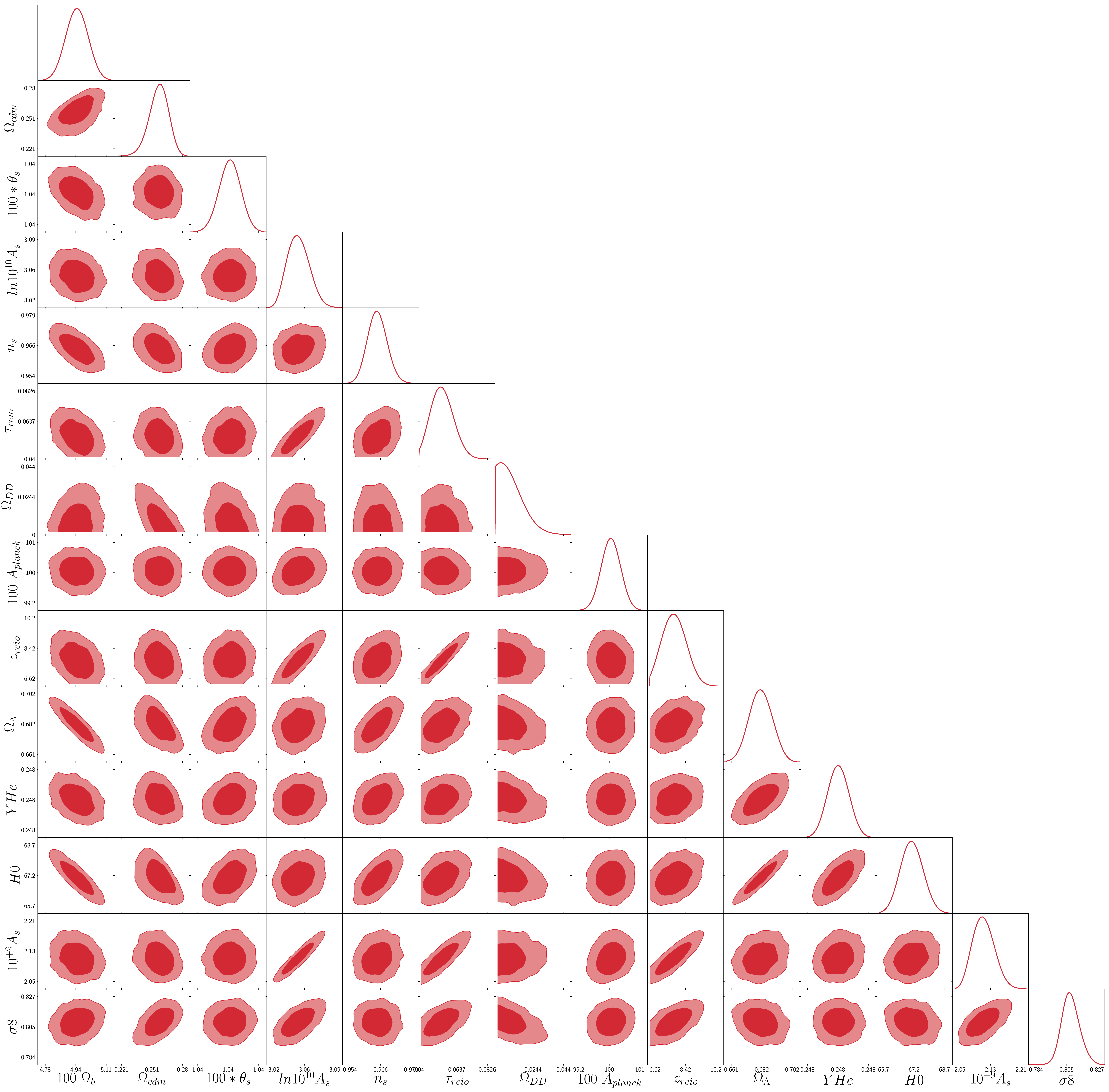}
\caption{Triangle graph from the MCMC analysis for the DD model is shown  for $m_{\rm dd}=10 \, \rm GeV$.}
\label{fig:dd_post}
\end{figure}

In Figures~\ref{fig:ccdm_post} and \ref{fig:dd_post},   we show the contour plots   of cosmological parameters. The estimated best-fit parameters and their 2-$\sigma$ errors are listed in Tables~\ref{T-CCDM} and~\ref{T-DD}.

The comparison of our theoretical predictions with data shows that the current Planck data
is consistent with no electromagnetically-interacting component of dark matter;    $\Omega_{\rm CCDM}$ and $\Omega_{\rm DD}$ are  consistent with zero. Figures~\ref{fig:ccdm_post} and~\ref{fig:dd_post} show that less than a few percent
of the total non-relativistic dark  matter (nearly 1\% for the CCDM model and  4\% for the DD model) could receive contribution from these components. Our results for the CCDM model are in general  agreement with other cosmological analyses of the model (e.g. \cite{Kovetz:2018zan}).

\begin{table}
\begin{center}
{\renewcommand{\arraystretch}{1.2}%
\begin{tabular}{|l|c|c|c|c|} 
 \hline 
Param & best-fit & mean$\pm\sigma$ & 95\% lower & 95\% upper \\ \hline 
$100~\Omega_{b }$ &$4.978$ & $4.916_{-0.064}^{+0.068}$ & $4.784$ & $5.05$ \\ 
$\Omega_{\rm cdm }$ &$0.2695$ & $0.2612_{-0.0076}^{+0.0084}$ & $0.2453$ & $0.2771$ \\ 
$100*\theta_{s }$ &$1.042$ & $1.042_{-0.00028}^{+0.00032}$ & $1.041$ & $1.042$ \\ 
$\rm ln10^{10}A_{s }$ &$3.045$ & $3.049_{-0.016}^{+0.014}$ & $3.021$ & $3.078$ \\ 
$n_{s }$ &$0.9625$ & $0.9663_{-0.0044}^{+0.0045}$ & $0.9573$ & $0.9754$ \\ 
$\tau_{\rm reio }$ &$0.05397$ & $0.05576_{-0.0081}^{+0.0068}$ & $0.04148$ & $0.07001$ \\ 
$\Omega_{\rm CCDM}$ &$0.0004058$ & $0.003395_{-0.0034}^{+0.00067}$ & $3.143\times 10^{-7}$ & $0.009859$ \\ 
$100~A_{\rm planck }$ &$99.95$ & $100.1_{-0.25}^{+0.26}$ & $99.53$ & $100.6$ \\ 
$z_{reio }$ &$7.686$ & $7.84_{-0.78}^{+0.71}$ & $6.362$ & $9.221$ \\ 
$\Omega_{\Lambda }$ &$0.6788$ & $0.6847_{-0.0078}^{+0.008}$ & $0.6694$ & $0.7002$ \\ 
$\rm YHe$ &$0.2478$ & $0.2478_{-6.7e-05}^{+7.3e-05}$ & $0.2477$ & $0.2479$ \\ 
$H0$ &$66.96$ & $67.34_{-0.58}^{+0.55}$ & $66.24$ & $68.44$ \\ 
$10^{+9}A_{s }$ &$2.1$ & $2.11_{-0.034}^{+0.029}$ & $2.052$ & $2.172$ \\ 
$\sigma_8$ &$0.8133$ & $0.8124_{-0.0063}^{+0.006}$ & $0.7999$ & $0.8245$ \\ 
\hline 
\end{tabular}} \quad \\

\caption{\label{T-CCDM} The best fit parameters and 2$\sigma$ errors
  from MCMC analysis are displayed  for the CCDM model, for the following
  parameters kept fixed: $m_{\rm ccdm} = 50 \, \rm MeV$ and $q_{\rm ccdm} = 10^{-6}$e. The remaining parameter, the fraction of the milli-charged dark matter, $f_{\rm ccdm}$ is varied,  yielding  the constraint on the density parameter  $\Omega_{\rm ccdm} \equiv f_{\rm ccdm} \Omega_{\rm dm}$.}

\end{center}
\end{table}

\begin{table}
\begin{center}
{\renewcommand{\arraystretch}{1.2}%
\begin{tabular}{|l|c|c|c|c|} 
 \hline 
Param & best-fit & mean$\pm\sigma$ & 95\% lower & 95\% upper \\ \hline 
$100~\Omega_{b }$ &$4.941$ & $4.952_{-0.06}^{+0.064}$ & $4.828$ & $5.074$ \\ 
$\Omega_{\rm cdm }$ &$0.2623$ & $0.2571_{-0.0087}^{+0.01}$ & $0.2377$ & $0.2759$ \\ 
$100*\theta_{s }$ &$1.042$ & $1.042_{-0.00029}^{+0.00032}$ & $1.041$ & $1.042$ \\ 
$\rm ln10^{10}A_{s }$ &$3.048$ & $3.049_{-0.014}^{+0.013}$ & $3.023$ & $3.076$ \\ 
$n_{s }$ &$0.9649$ & $0.965_{-0.0041}^{+0.0041}$ & $0.9568$ & $0.9729$ \\ 
$\tau_{\rm reio }$ &$0.05517$ & $0.0547_{-0.0078}^{+0.0062}$ & $0.04136$ & $0.06814$ \\ 
$\Omega_{\rm DD}$ &$0.004163$ & $0.01086_{-0.011}^{+0.0028}$ & $4.962\times 10^{-6}$ & $0.02694$ \\ 
$100~A_{\rm planck }$ &$100.1$ & $100.1_{-0.24}^{+0.25}$ & $99.55$ & $100.5$ \\ 
$z_{\rm reio }$ &$7.801$ & $7.746_{-0.76}^{+0.65}$ & $6.379$ & $9.105$ \\ 
$\Omega_{\Lambda }$ &$0.6826$ & $0.6811_{-0.0081}^{+0.0073}$ & $0.666$ & $0.6967$ \\ 
$\rm YHe$ &$0.2478$ & $0.2478_{-6.6e-05}^{+6.6e-05}$ & $0.2477$ & $0.2479$ \\ 
$H0$ &$67.18$ & $67.09_{-0.57}^{+0.54}$ & $66$ & $68.22$ \\ 
$10^{+9}A_{s }$ &$2.108$ & $2.11_{-0.031}^{+0.027}$ & $2.054$ & $2.167$ \\ 
$\sigma_8$ &$0.811$ & $0.8087_{-0.0065}^{+0.007}$ & $0.7953$ & $0.822$ \\ 
\hline 
 \end{tabular}} \quad \\ 

\caption{\label{T-DD} MCMC results for the DD model with $m_{\rm dd} = 10 \, \rm GeV$. $\Omega_{\rm dd} \equiv f_{\rm dd} \Omega_{\rm dm} $, the energy density in the form of DD atoms, is constrained by the CMB data}

\end{center}
\end{table}

\section{Conclusion and future prospects} \label{sec:disc}
We have studied cosmological implications of  three  models of cold dark matter in which the dark matter interacts electromagnetically.

Our paper has three main aims:
\begin{itemize}
\item To study the formation, stability, and interaction of two classes
  of dark matter atoms, DD and HeD (Section~\ref{sec:appendix}). The third model, the  extensively-studied  milli-charged dark matter (CCDM),  is used as a reference model for our study. 
\item To  understand and compute, both analytically and numerically, the cosmological implications of these models (Section~\ref{sec:tighcou}).
  \item To compare the predictions of these models with Planck CMB data.
\end{itemize}
One of the models (CCDM) 
involves coulomb interaction between charged particles and the dark matter. In this case, the interaction cross-section falls as $1/v^4$ which causes stronger
interaction at later times and therefore it impacts large scales which enter 
the horizon later. 
The other two models we study correspond to   neutral atoms formed by the recombination of    either two heavy singly  charged   particles (DD) or 
 the recombination of  a  heavy   doubly-charged   particle with Helium nucleus (HeD). For the  parameters of interest, the cross-section of interaction of these atoms
with baryons is independent of velocity, which means the interaction is stronger 
at early times and  the matter power spectrum at small scales is  affected. 

The DD and HeD models  are   qualitatively  different from each other  because 
the interaction cross-section  strongly depends on the mass of the dark matter particle in the  
DD case while it is independent of the dark matter mass for the  HeD atom. The HeD 
model is the most natural  extension  of the  dark matter paradigm   that admits a  stable, electromagnetically-interacting  dark matter particle. The inference follows as  the data rules out the other two compelling choices:  stable dark matter particle of electronic charge   or a neutral  atom  formed by the recombination of a  proton and a dark matter particle.  Our analysis shows that, for the range of  other parameters allowed by astrophysical data,  the HeD atom can  constitute all  the cold dark matter. 

In the recent past, dark matter models with electromagnetic interactions have
been invoked to explain the EDGES result \cite{EDGES2018,Barkana2018,fialkov18,Kovetz:2018zan}. The EDGES result requires the baryons
to be cooler at $z\simeq 20$ as compared to the usual model. One plausible explanation of this result is  coloumbic interactions between milli-charged  dark 
matter and baryons. It is of interest to investigate the cosmological implications 
of the other two models we study  in the post-recombination  universe. We note
that models of atomic dark matter cannot  explain the EDGES result
as, unlike the CCDM model,  their interaction with baryons are significant only in the early universe.

In this paper, we compare our theoretical predictions with Planck CMB data which 
is sensitive to Fourier modes, $k < 0.2 \, \rm Mpc^{-1}$ (e.g. \cite{Planck2018,Ade:2015xua}).
The DD model (Figure~\ref{fig:dd_pow}) also  allows for  significant difference 
in the matter power at smaller scales that cannot be  probed by  CMB data. The matter power at these scales can be constrained  by  cosmological observables such as weak lensing
and the clustering of Lyman-$\alpha$ clouds (e.g. \cite{Sarkar:2014bca}) and other data such as Milky way satellite population (e.g. \cite{Nadler_2019}).  We hope to return to this 
comparison in the future. 

If the dark matter particle interacts  electromagnetically, it might be 
easier to detect it directly and it might have a bearing on the formation
of stars and  galaxies (see e.g. \cite{2013PhRvD..87j3515C,1990NuPhB.333..173D,1990PhRvD..41.2388D,fialkov18}). The constraints on the CCDM model from 
such considerations have been discussed extensively (e.g. \cite{fialkov18}), 
we briefly discuss here the impact of atomic dark matter on the formation of 
galaxies. The observed spiral galaxies display baryonic component in a disk
surrounded by a halo whose mass is dominated by dark matter. The separation of 
baryons from dark matter occurs because baryons can cool and therefore  fall towards the center of the potential well of a  virialized halo while the dark matter dominates the halo mass. The main cooling mechanisms for haloes (of primordial chemical  composition)  in the temperature range of $10^4\hbox{--}10^6 \, \rm K$ are  line de-excitation of neutral hydrogen, line de-excitation of  singly-ionized helium, and free-free emission (see e.g. \cite{2011piim.book.....D}). For a dark matter interacting electromagnetically with baryons, we might expect dark matter to also cool and  fall to the center of potential wells, which would be in contradiction with  the observables  of galaxies. Therefore, it is important to show
that the dark matter atoms, and in particular the HeD and DD atoms which could constitute all the cold dark matter, do not behave as baryons for the energy scales and 
densities of interest in a galaxy ($T \simeq 10^5 \, \rm K$ and $n_b \simeq 0.01 \,\rm cm^{-3}$). First, these  atoms cannot cool owing to any of the mechanisms discussed above because their  energy scales (binding energy  and atomic levels) are too large to be affected by  collisions with baryons  which could cause line de-excitation or ionization. Second, the time scale of collisions of  HeD or DD atoms 
with baryons   is generally larger than the age of the 
universe   (Eqs.~(\ref{eq:ddhe_intrate}) and~(\ref{eq:inttime})). This implies that HeD and  DD atoms behave like a non-interacting dark matter particle for typical energy scales  and densities
expected in a galaxy. It is possible that these particles could interact 
with baryons and photons in denser parts of the galaxies\footnote{The additional
interaction might help in resolving  the  core-cusp and ``too big to fail''  issues as one possible way to address these discrepancies is to invoke  dark matter self-interaction, e.g.  see \cite{2018PhR...730....1T,Foot_2016} and references therein.}  or where the energy 
scales are larger (e.g. supernova remnants).

The aim of this paper is to investigate the cosmological implications
of a dark matter particle with electromagnetic interactions. It is  possible
the parameter space we studied could  be further constrained from
other astrophysical and experimental constraints.  We leave such an investigation to a future work. 

\acknowledgments
{
Our results are based on observations obtained with Planck (http://www.esa.int/Planck), an ESA science mission with instruments and contributions directly funded by ESA Member States, NASA, and Canada.
}

\newpage

\bibliographystyle{plain}
\bibliography{DRAFT_fin}
 
\end{document}